\documentclass[11pt]{article}
\pdfoutput=1
\usepackage{jheppub}

\usepackage[utf8]{inputenc}
\usepackage{booktabs}
\usepackage[dvipsnames]{xcolor}
\usepackage[capitalise]{cleveref}
\usepackage{siunitx}
\DeclareSIUnit{\year}{yr}
\DeclareSIUnit{\parsec}{pc}
\usepackage[normalem]{ulem}

\newcommand{\refcite}[1]{Ref.~\cite{#1}}
\newcommand{\refscite}[1]{Refs.~\cite{#1}}
\usepackage{aas_macros}

\usepackage{bm}
\newcommand{\bb}[1]{\bm{\mathrm{#1}}}
\newcommand{\du}{\mathrm{d}}
\newcommand{\dd}{\,\du}
\newcommand{\dm}{\mathrm{DM}}
\newcommand{\argmin}{\operatorname{argmin}}
\renewcommand{\Re}{\operatorname{Re}}
\renewcommand{\Im}{\operatorname{Im}}
\newcommand\erf{\operatorname{erf}}
\newcommand{\host}{\mathrm{h}}
\newcommand{\soliton}{\mathrm{s}}
\newcommand{\eff}{\mathrm{eff}}
\newcommand\tidal{\mathrm{tidal}}
\newcommand\core{\mathrm{c}}
\newcommand\euc{\mathrm{E}}

\begin{document}
\title{Tunneling and tidal stripping in multifield ultralight dark matter halos}

\preprint{%
    \parbox[t]{4.5cm}{\raggedleft
        IFT-UAM/CSIC-26-60\\
        MIT-CTP/5961}}
\date\today

\author[a]{Benjamin V. Lehmann,}
\affiliation[a]{Center for Theoretical Physics---a Leinweber Institute, Massachusetts Institute of Technology, Cambridge, MA 02139, USA}

\author[b]{Jackie Lodman,}
\affiliation[b]{Department of Physics, Harvard University, Cambridge, MA 02138, USA}

\author[c]{and Thomas Steingasser}
\affiliation[c]{Departamento de Fisica Te\'orica, Universidad Autonoma de Madrid, and IFT-UAM/CSIC, Cantoblanco, 28049, Madrid, Spain}

\abstract{\ignorespaces
    Tidal stripping is a key feature of the evolution of dark matter (DM) halos, and has major implications for the population of low-mass galaxies. In the case of ultralight DM, tidal stripping proceeds not only classically, at the tidal radius, but also via a process analogous to quantum tunneling by long-wavelength particles out of the potential of a subhalo. This modified tidal stripping behavior leads to tight constraints on the particle mass as a function of subhalo and host properties. As many models of ultralight DM predict several independent species, it is crucial to understand how these constraints can be generalized to multifield halos with different particle masses. However, numerical challenges make it difficult to directly study the tunneling process in all but the simplest multifield scenarios. We introduce a simplified approach based on semiclassical methods that entirely sidesteps the most difficult aspects of the numerical problem, and we apply this to the study of tunneling in multifield halos. Our results significantly clarify the physics of tidal stripping for ultralight DM halos even in the single-field case: we provide first-principles derivations of features of the tunneling rate previously suggested by empirical fits. We then evaluate stability bounds on two-field halos for the first time, for a wide range of density and particle mass ratios. We show that for particular parameter combinations, the stability bounds in the two-field case can be somewhat relaxed relative to the single-field case, but for much of the parameter space, the constraints become more stringent. We discuss the path towards probing realistic multifield ultralight DM halos.
}


\maketitle

\section{Introduction}
While the cosmological abundance of dark matter (DM) is well measured, some of its most basic properties are understood little better now than at the dawn of DM science~\cite{deSwart:2017heh,Peebles:2022bya}. In particular, the mass of the DM particle may lie anywhere within an enormous range, spanning roughly eighty orders of magnitude~\cite{Bertone:2004pz,Profumo:2017hqp,Cirelli:2024ssz}. Setting even these bounds has been the subject of an enormous effort, beginning from the observation that the mass of the DM has an impact on the properties of the collapsed structures that it can form. At the highest masses, dynamical bounds on massive compact objects as DM arguably provided the first-ever limit on the properties of DM~\cite{1969Natur.224..891V}, and have continued to provide leading upper bounds on the mass of the DM constituent~\cite{Brandt:2016aco,Penarrubia:2016ltr,Koushiappas:2017chw,Graham:2023unf}. Similarly, at the lowest masses, the extended wavelength of ultralight DM particles leads to constraints from the existence of small-scale DM structures~\cite{Hu:2000ke,Schive:2014dra,Hui:2016ltb,Irsic:2017yje,Hui:2021tkt}. 

While setting the strongest bounds requires very careful work, the rough scales of the resulting limits are easy to understand, and follow directly from the properties of observable DM structures. A small galaxy has a mass of order \qty{e8}{M_\odot}, and a spatial extent of order \qty{1}{\kilo\parsec}. Thus, a DM particle should have mass and spatial extent smaller than each of these quantities. The spatial extent of a DM particle can be estimated by its de Broglie wavelength $\lambda_{\mathrm{dB}}$, and $\lambda_{\mathrm{dB}} \lesssim \qty{1}{\kilo\parsec}$ implies a bound $m_\dm \gtrsim \qty{e-23}{\electronvolt}$. State-of-the-art bounds on the DM mass are several orders of magnitude stronger than these estimates on either side, requiring $\qty{e-20}{\electronvolt}\lesssim m_\dm \lesssim \qty{100}{M_\odot}$\footnote{This upper limit is the bound from dynamics alone. In practice, it is generally agreed that a combination of bounds from accretion, lensing, and gravitational waves further limit the maximum DM mass to about \qty{e-13}{M_\odot}, or \qty{e53}{\electronvolt}~\cite{Green:2020jor,Carr:2020gox,Carr:2020xqk}.}~\cite{Rogers:2020ltq,Tyler:2022rxi,Graham:2023unf}.

Inevitably, DM masses near either of these bounds lead to nontrivial effects that are not quite strong enough to rule out such parameter values, but still significant enough to induce a range of interesting phenomena potentially accessible to observations. Accordingly, both ultramassive and ultralight DM have been invoked to account for tensions between data and the small-scale predictions of $\Lambda$CDM\@. Ultralight DM in particular was thought to be a simple and well-motivated solution to these small-scale ``crises.'' A deficiency of small galaxies (i.e., the missing satellites problem~\cite{Hu:2000ke, Weinberg:2013aya}) and the unexpected formation of cores in the density profiles of larger galaxies (i.e., the core-cusp problem~\cite{Hui:2016ltb,Weinberg:2013aya,DelPopolo:2021bom}) are both readily explained if the DM particle is \emph{somewhat} wavelike: compact enough to admit the smallest observed structures, but extended enough to suppress the formation of small galaxies and cusp-like features at galactic centers. At the same time, the existence of an ultralight species such as a light axion or dilaton is independently motivated by string-theoretic constructions~\cite{Armengaud:2017nkf,Rogers:2020ltq,Kobayashi:2017jcf}.

The combination of these two factors led to great interest in ultralight DM with a mass between \num{e-23} and \qty{e-21}{\electronvolt}~\cite{Hu:2000ke}. In addition to suppressing small-scale structure formation, ultralight DM naturally explains the presence of constant-density cores in galaxies. Wavelike DM, when relaxed into its quantum mechanical ground state, forms a solitonic structure whose density profile features exactly such a core. Thus, beyond ameliorating small-scale problems, ultralight DM offers the exciting opportunity to observe macroscopic quantum effects on the scale of galaxies. These effects exhibit a rich phenomenology that has given rise to a number of new observational signatures. For example, interference fringes result in dense regions within the galactic halo that behave much like compact objects~\cite{Schive:2014dra}.

Moreover, the standard physical effects that govern the assembly and evolution of halos in $\Lambda$CDM can be substantially modified by the long wavelength of ultralight DM. A key example is tidal stripping, which governs the survival of satellite galaxies in $\Lambda$CDM\@. In general, a galaxy embedded in the potential of a larger host halo is subjected to tidal forces that can strip away DM, with significant implications for the population of low-mass galaxies. In ultralight DM halos, tidal stripping can also proceed by a process resembling quantum tunneling: the gravity of the satellite halo creates a potential barrier that keeps DM bound to the satellite, but in the presence of a tidal potential, the DM can tunnel through the barrier and depart from the subhalo. This process predicts a set of striking relationships between the properties of satellites and hosts that are directly testable~\cite{Hui:2016ltb,Hertzberg:2022vhk, Dave:2023egr}.

Over the last ten years, the relevance of small-scale structure tensions to DM physics has been called into question: the core-cusp problem may be resolved by the incorporation of baryonic physics into high-resolution simulations~\cite{Chan:2015tna,Onorbe:2015ija,Genina:2017gzv,Hopkins:2017ycn,Sales:2022ich}, and the missing satellites problem has been convincingly resolved with new datasets~\cite{DES:2020fxi} (although the so-called galaxy diversity problem arguably remains). At the same time, lower bounds on the DM mass have improved, excluding the minimal scenarios with $\mathcal O(\qty{e-21}{\electronvolt})$ masses originally invoked to account for small-scale tensions~\cite{Armengaud:2017nkf,Rogers:2020ltq,Kobayashi:2017jcf,Pozo:2023zmx, Hayashi:2021xxu, Safarzadeh:2019sre, Dalal:2022rmp}. But the UV motivation for ultralight species is still strong, and simultaneously suggests conditions which violate the assumptions of most lower bounds on the mass. In particular, UV constructions readily allow ultralight DM to consist of many mutually- and self-interacting species. The presence of multiple species and the presence of interactions both greatly influence the ground state into which the DM particles relax, modifying the structure of the halo, and hence the constraints that can be placed on the DM mass.

Studying these effects, however, poses a significant challenge. For particle-like DM (i.e., DM with a small spatial extent), many important features of halo structure and evolution have been extracted from numerical simulations that are built on treating the DM as a gravitationally-interacting fluid. Similar techniques can be used to solve for the evolution of DM in the wavelike limit, but the DM distribution must be described with a wave function evolving under the Schr\"odinger equation, satisfying a different set of evolution equations from a particle-like DM fluid. Relatively few simulations of this kind have been performed in the wavelike limit. Soliton structure and tidal stripping have been studied in 3d simulations by \refcite{Du:2018qor} for a single DM species. Solitons with self-interactions and a combination of two species have only recently been simulated by the authors of \refscite{Luu:2018afg, Huang:2022ffc,Luu:2024gnk}. Given the paucity of simulation suites, a full picture of halo evolution from simulations is still lacking.

On the other hand, under the ansatz that an ultralight DM halo relaxes to its quantum mechanical ground state, a host of techniques from quantum mechanics can be employed to determine at least some of the key features of the DM halo and its evolution. The structure of an isolated soliton can be obtained directly by solving the time-independent Schr\"odinger equation for the ground state under the assumption of spherical symmetry, via a 1d differential system~\cite{Hui:2016ltb, Marsh:2015wka}. Tidal stripping, while inherently asymmetric and time-dependent, can also be quantified in the same framework by treating the ground state as an approximate energy eigenstate---i.e., a resonance or quasistationary state---with a complex energy eigenvalue. The imaginary part of the eigenvalue gives the decay rate of the ground state, which estimates the rate of depletion under tidal stripping by a process resembling quantum tunneling. While approximate, this computation is sufficient to establish key features of satellite--host interactions in ultralight DM halos~\cite{Hui:2016ltb,Hui:2021tkt}.

One would hope that multifield halos could be studied by similar means, by solving for the quasistationary ground state of a system with two or more fields together. However, the extension of these techniques to the case of multiple DM fields has proven to be quite nontrivial~\cite{Luu:2018afg, Luu:2023dmi}. Solving for the approximate ground state requires numerical solution of an eigenvalue problem by a shooting method, and extreme numerical precision is required for many parameter regions of interest, even for single-field solitons. For a two-field halo, the analogous shooting problem is not only even more demanding in numerical precision, but four-dimensional rather than two-dimensional, as each field admits its own complex eigenvalue. Thus, some of the most interesting cases, such as scenarios with a large ($\mathcal O(10)$) mass ratio between the two fields, or with very mild tidal forces, have remained difficult to study.

At the same time, exploring such scenarios is crucial in order to resolve the status of constraints on ultralight DM\@. Multifield halos are potentially compatible with observables where single-field ultralight DM is robustly excluded, since the constraints arise in part from comparisons between different systems, so allowing for different mixtures of two or more fields between these systems immediately relaxes constraints. In particular, \refscite{Luu:2018afg,Pozo:2023zmx} found approximate numerical solutions for the ground state with two DM species, and showed explicitly that mixed halos are compatible with data even at very low masses if the two species have a mass ratio of $\mathcal O(10)$.

A considerable literature has thus developed in pursuit of the general features of multifield halos. Several authors have studied the conditions under which multifield solitons can form~\cite{Huang:2022ffc, vanDissel:2023vhu, Luu:2023dmi, Luu:2024gnk}, the extent to which multifield solitons can be treated as a combination of two isolated solitons~\cite{Luu:2018afg, Luu:2023dmi}, and the degree to which constraints on single-field DM can be enhanced~\cite{Luu:2018afg,Gosenca:2023yjc}. But basic questions about the physics of multifield halos still remain unanswered, particularly with regard to tidal stripping. Since tidal stripping in ultralight DM halos receives contributions from the quantum tunneling process, it is not obvious how the rate of tidal stripping is modified in the presence of multiple fields, nor is it easy to extract such tunneling rates by identifying complex eigenvalues as for the single-field case.

In this work, we introduce new methods to compute tidal stripping rates for multifield halos. Specifically, we first develop a numerical technique capable of reliably computing the real parts of the complex ground-state eigenvalues, even with large mass ratios. We then argue that for all cases of physical interest, the real parts alone are sufficient to exhibit the most important features of a two-field soliton. In particular, the rate of tidal stripping, historically computed using the imaginary parts of the eigenvalues, can be readily obtained from semiclassical methods such as the WKB approximation and instanton techniques once the real parts are known. We use this framework to study tidal stripping in multifield halos, rapidly scanning over parameter space in a fashion inaccessible to expensive simulations.

The remainder of this work is organized as follows. In \cref{sec:single-field}, we review the usual treatment of spherically-symmetric single-field solitons, and the computation of tidal stripping rates in the quasistationary formalism (i.e., from the complex eigenvalue). In \cref{sec:two-field}, we study two-field solitons in the same framework. We describe the numerical challenges that prevent the application of this method to arbitrary configurations, and we establish benchmarks from particular limits in which the two-field system is readily solved. Then, in \cref{sec:numerics}, we describe our numerical procedure for determining only the real part of each eigenvalue, and we show that this is sufficient to recover the basic structure of the two-field soliton. In \cref{sec:semiclassics}, we introduce semiclassical methods to compute the decay rates themselves, and apply this to compute subhalo lifetimes in \cref{sec:multifield-lifetimes}. We discuss applications of our results and conclude in \cref{sec:conclusions}.

In much of this work, we use nondimensionalized and rescaled variables. We generally write dimensionful variables with a tilde (\,$\tilde\cdot$\,) and nondimensionalized variables with a hat symbol (\,$\hat\cdot$\,), reserving unadorned symbols for variables that are both nondimensionalized and rescaled. However, apart from these nondimensionalizations, we do \emph{not} use natural units, preserving explicit factors of $\hbar$ and $c$ to maximize the clarity of our definitions.

\section{Solitons and tidal stripping in single-field halos}
\label{sec:single-field}
We begin by reviewing the treatment of ultralight DM solitons in the single-field case in some detail, following \refscite{Hui:2016ltb, Marsh:2015wka, Chakrabarti:2022owq, Dave:2023egr}. For simplicity, we will work with nondimensionalized variables in much of our treatment. We thus use an overset tilde (\,$\tilde\cdot$\,) to denote relatively rare instances of \emph{dimensionful} variables. Note that this is opposite to the conventions of some other works in the literature, e.g. \refcite{Dave:2023egr}.

We begin from the classical action of a scalar field $\tilde{\Phi}$ with mass $\tilde m$ minimally coupled to gravity, which has the form
\begin{equation}
    \tilde S =
    \int\du^4\tilde{x}\,
    \sqrt{-\tilde{g}}
    \left(
        \tfrac12 M^2_\mathrm{Pl} \tilde{\mathcal{R}}
        - \tfrac12\tilde{\partial}^\mu\tilde{\Phi}
            \tilde{\partial}_\mu \tilde{\Phi}-\tfrac12\tilde {m}^2\tilde{\Phi}^2
        - \tilde{U}(\tilde{\Phi})
    \right)
    ,
\end{equation}
for some interaction potential $\tilde{U}(\tilde{\Phi})$, where $\tilde{\mathcal{R}}$ denotes the Ricci scalar. In the weak-gravity regime, the line element is given by
\begin{equation}
    \du\tilde{s}^2 =
    - \bigl(1+2\tilde{V}\bigr) \dd{{\tilde t}\,}^2
    + \bigl(1-2\tilde{V}\bigr) \dd{\tilde{\bb x}}^2,
\end{equation}
where $\tilde V$ denotes the gravitational potential. In the non-relativistic limit, we can separate $\tilde\Phi$ into rapidly-oscillating exponential factors and a slowly-varying component $\tilde\phi(\tilde t, \tilde{\bb x})$, as
\begin{equation}
    \tilde{\Phi} =
    \frac{\hbar}{\sqrt{2}\tilde m}\left(
        e^{-i\tilde mc^2\tilde{t}/\hbar}\tilde{\phi}(\tilde{t},\tilde{\bb x})
        + e^{i\tilde mc^2\tilde t/\hbar}\tilde{\phi}^*(\tilde{t},\tilde{\bb x})
    \right)
    .
\end{equation}
The rapidly-oscillating components can be neglected in time averages, so $\langle|\tilde\Phi|^2\rangle \simeq (\hbar^2/\tilde m)|\tilde\phi|^2$. The equation of motion for $\tilde\Phi$, along with Einstein's equations for a perfect fluid, then yield the \textit{Gross--Pitaevskii--Poisson system} for $\tilde\phi$:
\begin{equation}
    \label{eq:single-field-time-dependent}
    i\hbar \frac{\partial\tilde{\phi}}{\partial \tilde t} =
    - \frac{\hbar^2}{2\tilde m}\tilde \nabla^2\tilde{\phi}
    + \tilde{m}\tilde{V}_{\eff}\tilde{\phi},
    \qquad
    \tilde\nabla^2 \tilde{V}=4\pi G|\tilde{\phi}|^2,
\end{equation}
where $\tilde{V}_{\eff}$ is the full effective potential. In general, $\tilde V_{\eff} = \tilde{V} + \tilde{V}_{\mathrm{int}}$ for some interaction potential $\tilde{V}_{\mathrm{int}}$. In the limit $\tilde{V}_{\mathrm{int}}\rightarrow0$, we expect stationary solutions of the form 
\begin{equation}
    \label{eq:free-form}
    \tilde{\phi}(\tilde{t},\tilde{\bb x}) =
    \tilde{\chi}(\tilde{\bb x})\,e^{-i\tilde{\gamma} \tilde{t}/\hbar}
    \quad \text{for~}\tilde{\gamma} \in \mathbb{R}.
\end{equation}
This is analogous to the usual decomposition of a solution of the time-dependent Schr\"odinger equation into a time-independent part and an exponential time-evolution factor. Here $\tilde\gamma$ plays the role of the energy eigenvalue, and for a perfectly stationary solution, $\tilde\gamma$ is real. However, if $\tilde{V}_{\mathrm{int}}\neq0$, we generally expect nonstationary solutions, meaning that $\tilde\gamma$ acquires an imaginary part, and $e^{-i\tilde{\gamma} \tilde{t}/\hbar}$ is no longer a pure phase. We will be interested in quasistationary solutions, where $\Im(\tilde{\gamma}) \ll \Re(\tilde{\gamma})$ for small $\tilde V_{\mathrm{int}}$, and the parametrization of \cref{eq:free-form} is still useful.

We thus rewrite \cref{eq:single-field-time-dependent} in terms of $\tilde\chi$, as
\begin{equation}
    \label{eq:free-system}
    \tilde\gamma\tilde{\chi} =
    -\frac{\hbar^2}{2\tilde m}\tilde{\nabla}^2\tilde{\chi}
    +\tilde m \tilde{V}_{\eff}\tilde{\chi},
    \qquad
    \tilde\nabla^2 \tilde{V} = 4\pi G|\tilde{\chi}|^2.
\end{equation}
For actual computations, it is useful to rewrite this system with a set of dimensionless variables, which we denote with hats (\,$\hat\cdot$\,), defined as follows:
\begin{equation}
    \label{eq:dimensionless-variables} 
    \hat{\chi}=\frac{\hbar\sqrt{4\pi G}}{\tilde mc^2}\tilde{\chi},
    \qquad
    \hat{V}=\frac{\tilde{V}}{c^2},
    \qquad
    \hat{\gamma}=\frac{\tilde{\gamma}}{\tilde mc^2},
    \qquad
    \hat{\bb x}=\frac{\tilde mc}{\hbar}\tilde{\bb x},
    \qquad
    \hat{t}=\frac{\tilde mc^2}{\hbar}\tilde{t}
    .
\end{equation}
The nondimensionalization of other quantities can be readily inferred from these. In these variables, \cref{eq:free-system} simplifies to
\begin{equation}
    \label{eq:free-system-dimensionless}
    \hat{\gamma}\hat{\chi} =
    -\frac{1}{2}\hat{\nabla}^2\hat{\chi}
    +\hat{V}_{\eff}\hat{\chi},
    \qquad
    \hat{\nabla}^2 \hat{V}=|\hat{\chi}|^2.
\end{equation}
We now seek (quasi)stationary solutions of this system, so we treat $\hat\chi$ as a function of $\hat{\bb x}$ alone. We further specialize to spherical symmetry, so that $\hat\chi$ is a function only of the radial coordinate $\hat r$, measured from the center of the halo under consideration. The boundary conditions on this system are subject to the properties of a given DM halo. However, for a halo with a central density $\tilde\rho_\core$, the following boundary conditions apply:~\cite{Marsh:2015wka}
\begin{equation}
    \hat{\chi}(0)=\sqrt{\hat\rho_\core},
    \qquad
    \hat{\chi}'(0)=0,
    \qquad
    \hat{\chi}(\infty)=0,
    \qquad
    \hat{V}'(0)=0,
\end{equation}
where $\hat\rho_\core$ is the nondimensionalized DM density. We further choose to set the potential to zero at the origin, i.e., $\hat{V}(0)=0$. (This is physically inconsequential, but note that it does affect the definition of the energy eigenvalue $\hat\gamma$, since a shift in one of the two can be compensated by a shift in the other.)

Up to this point, we have left the form of $V_{\eff}$ completely generic. In this work, we are interested in tidal stripping, which takes place via a \textit{tidal potential.} This is the contribution to the effective potential that arises due to the presence of a spatially-varying external potential. In particular, we take our system to be a subhalo orbiting the center of a host halo with mass $\tilde M_\host$ at a distance $\tilde R$ with angular velocity $\tilde\omega$, or period $\tilde T = 2\pi/\tilde\omega$. We assume that the subhalo has a mass $\tilde M_\soliton \ll \tilde M_\host$. As long as the distance from the center of the subhalo, $\tilde r$, satisfies $\tilde r \ll \tilde R$, the tidal potential can be expanded along the axis separating the two halo centers, as
\begin{equation}
    \tilde V_\tidal =
    -\frac{3}{2}\frac{G\tilde M_\host}{\tilde R}
        \left(\frac{\tilde r}{\tilde R}\right)^2
    + \mathcal O\left[\left(\frac{\tilde r}{\tilde R}\right)^3\right]
    .
\end{equation}
Kepler's third law gives $\tilde\omega^2 = G\tilde M_\host/\tilde R^3$, so we can write $\tilde V_\tidal(\tilde r) \approx -\frac32\tilde\omega^2\tilde r^2$. For the remainder of this work, we set $\tilde V_\eff = \tilde V + \tilde V_\tidal$, and assume the form above for $\tilde V_\tidal$. That is, we take
\begin{equation}
    \tilde V_\eff = \tilde V - \tfrac32\tilde\omega^2\tilde r^2
    .
\end{equation}
The angular velocity $\tilde\omega$ is thus the crucial parameter for tidal stripping: it is the single parameter that fixes the tidal potential.

Given this form for $\tilde V_\eff$ and its nondimensionalization $\hat V_\eff$, we can further simplify the boundary conditions by introducing a rescaling. The system of \cref{eq:free-system-dimensionless} is invariant under the following transformation, for any $s$:
\begin{equation}
    \hat t \to s^2 \hat t,
    \quad
    \hat r \to s \hat r,
    \quad
    \hat\chi \to s^{-2} \hat\chi,
    \quad
    \hat V \to s^{-2} \hat V,
    \quad
    \hat\gamma \to s^{-2} \hat\gamma,
    \quad
    \hat\omega \to s^{-2} \hat\omega,
\end{equation}
where the nondimensionalization $\hat\omega$ is given by $\tilde{\omega}\hbar/\tilde m c^2$, following the conventions of \cref{eq:dimensionless-variables}. We choose the scaling factor $s$ so that $\hat\chi(0)$ is rescaled to 1. For a central density $\tilde\rho_\core$, this corresponds to
\begin{equation}
    \label{eq:scaling-factor}
    s(\tilde\rho_\core) = \left(
        \frac{\tilde m^2c^4}{4\pi G \hbar^2 \tilde{\rho}_\core}
    \right)^{1/4}
    .
\end{equation}
Once we impose this choice of scaling, we write the \emph{rescaled, nondimensionalized} variables without either tildes or hats---so, our boundary condition for $\chi$ is summarized by simply $\chi(0) = 1$. Our 1d system now takes the form 
\begin{equation}
    \label{eq:tidal-system}
    2\chi'(r)/r + \chi''(r) =
    2\left[V(r) - \tfrac32\omega^2r^2 - \gamma\right]\chi(r),
    \qquad
    2V'(r)/r + V''(r) = \left|\chi(r)\right|^2,
\end{equation}
with boundary conditions
\begin{equation}
    \chi(0)=1,
    \quad
    \chi'(0)=0,
    \quad
    \chi(\infty)=0,
    \quad
    V(0) = 0,
    \quad
    V'(0)=0.
\end{equation}
In practice, this is best treated numerically by separating the first equation of \cref{eq:tidal-system} into separate equations for its real and imaginary parts, as
\begin{equation}
    \label{eq:tidal-system-re-im}
    \begin{array}{l}
        (2/r)\partial_r\Re(\chi) + \partial_r^2\Re(\chi) =
            2\left[V(r) - \Re(\gamma)
                - \tfrac32\omega^2r^2\right]\Re(\chi)
            + 2\Im(\gamma)\Im(\chi)
        ,
        \\[2mm]
        (2/r)\partial_r\Im(\chi) + \partial_r^2\Im(\chi) =
            2\left[V(r) - \Re(\gamma)
                - \tfrac32\omega^2r^2\right]\Im(\chi)
            - 2\Im(\gamma)\Re(\chi)
        .
    \end{array}
\end{equation}

The value of $\gamma$ corresponds to the energy eigenvalue, and its value in the ground state must be determined from this system. In practice, imposing a boundary condition at infinity is numerically challenging, and two strategies are available. One approach is to compactify the system by introducing a coordinate of the form $u \propto \arctan(r)$. Alternatively, an approximate boundary condition at finite $r$ can be derived using the WKB approximation: in the simple scenario with no tidal potential ($\omega = 0$), in the large-$r$ regime, we may assume $V(r) \ll\gamma$ and $\gamma \in \mathbb R$, so we obtain $\chi'(r) / \chi(r) = -2\gamma$, i.e., an exponentially-decaying solution. We then simply seek solutions for which $\chi'(r) / \chi(r)$ takes a constant value at large $r$. If $\omega > 0$, on the other hand, then the tidal potential dominates in the large-$r$ regime, and the WKB solution implies that~\cite{Hui:2016ltb, Dave:2023egr}
\begin{equation}
    \chi(r) \longrightarrow
    \frac{A}{r^{3/2}}\exp\left(i\frac{\sqrt{3}}{2}\omega r^2\right)
\end{equation}
for some $A\in\mathbb C$. The boundary condition at infinity can then be implemented by matching onto a solution of this form at finite $r$. While these methods are numerically distinct, they are physically equivalent, and we refer to all such methods as \textit{boundary condition at infinity} (BCI) methods in the remainder of this work. We use the term \textit{quasistationary} for methods which solve directly for a complex energy eigenvalue. Thus, for $\omega > 0$, we refer to the procedure described above as the quasistationary BCI method. Numerically, this method can be challenging even for single-field halos, and we will see that it becomes intractable for two-field halos.

\begin{figure}
    \centering
    \includegraphics[width=\linewidth]{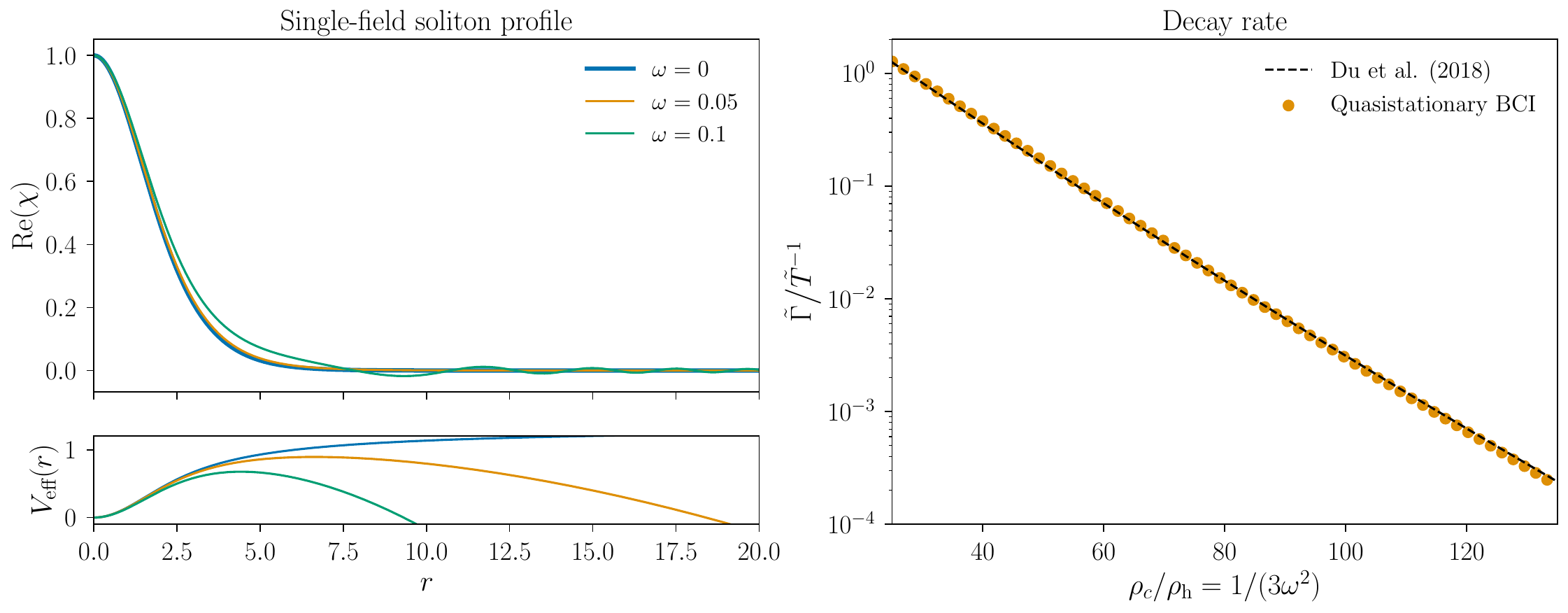}
    \caption{\textbf{Left:} $\Re(\chi)$ and $V_{\eff}$ for $\omega=0$ (blue), $\omega=0.05$ (orange), and $\omega=0.1$ (green). Note that we set $V(0)=0$. For $\omega>0$, the minimum at $r=0$ is a false vacuum, leading to $\Im(\gamma)\neq 0$ and $\Im(\chi)\neq 0$. The coupling between $\Re(\chi)$ and $\Im(\chi)$ then leads to the oscillations visible in the former.
    \textbf{Right:} The decay rate in units of $\tilde T^{-1}$ as a function of $\rho_\core/\rho_\host$. The decay rate increases exponentially with $\omega^2$, which sets the strength of the tidal potential. The dashed black curve shows the fitting function from~\refcite{Du:2018qor}.}
    \label{fig:single-field}
\end{figure}

The left panel of \cref{fig:single-field} shows numerical solutions of this system obtained with the BCI method for several different values of $\omega$. In the complete absence of the tidal potential ($\omega=0$), the minimum of the potential at $r=0$, corresponding to the center of the subhalo, is the global minimum. In this case, $\chi(r)$ and $\gamma$ are both fully real. However, for $\omega>0$, the center of the subhalo is only a local minimum of the potential: at large $r$, the tidal potential dominates, and drives $V_\eff(r) < 0$. Such a potential does not admit stable bound states, meaning that $\gamma$ (and thus $\chi$) must acquire an imaginary part. The imaginary part of $\chi$ is oscillatory, and its coupling to the real part (\cref{eq:tidal-system-re-im}) produces the oscillations visible in \cref{fig:single-field}. For later utility, we note that the single-field profile $\chi(r)$ at $\omega=0$ is well approximated by
\begin{equation}
    \label{eq:single-field-profile}
    \chi(r) \approx \exp(-0.1815r^2).
\end{equation}

Once we have obtained $\chi(r)$ and $\gamma$, the DM density profile, $\tilde\rho(\tilde r) = |\tilde\chi(\tilde r)|^2,$ can be used to determine the mass enclosed by a sphere of radius $\tilde r_*$. Using \cref{eq:free-form} to translate from $\chi$ to $\tilde\Phi$, we obtain
\begin{equation}
    \tilde M(\tilde t, \tilde r_*) =4 \pi 
        \int_0^{r_*}\du\tilde r\,\tilde r^2\,|\tilde\Phi(\tilde t,\tilde r)|^2
    = e^{2\Im(\gamma)t} \cdot 4 \pi \int_0^{r_*}\du\tilde r\,\tilde r^2\,|\tilde\chi(\tilde r)|^2
    .
\end{equation}
Accordingly, if $\Im(\gamma) < 0$, the mass of the ground state will decay with time. This is the contribution of tunneling to tidal stripping. In particular, it follows that the mass loss rate $\Gamma$ is related to the eigenvalue $\gamma$ by
\begin{equation}
    \label{eq:Gamma}
    \Gamma \equiv -\frac{1}{M}\frac{\du M}{\du t} = -2\Im{(\gamma)}.
\end{equation}
In the remainder of this work, we will often need to refer to the real and imaginary parts of $\gamma$ separately. Accordingly, we define $E = \Re(\gamma)$, so that
\begin{equation}
    \gamma \equiv E - \tfrac i2\Gamma
    .
\end{equation}
Restoring dimensions to $\Gamma$ gives the physical decay rate, $\tilde{\Gamma}=\tilde mc^2\Gamma/\hbar s^2$, where $s$ is the rescaling factor from \cref{eq:scaling-factor} used to set $\chi(0)=1$. Substituting, we obtain 
\begin{equation}
    \tilde\Gamma = \sqrt{4\pi G\tilde\rho_\core}\,\Gamma
    .
\end{equation}

Notably, all dependence on the DM mass has canceled: $\tilde m$ is absorbed into the scaling factor $s(\tilde\rho_\core)$, and the only dependence of the tidal stripping rate on the mass now appears through $\tilde\rho_\core$ itself. As discussed by \refscite{Du:2018qor,Hertzberg:2022vhk,Dave:2023egr}, the decay rate in this case depends exclusively on the ratio of the central density of the subhalo, $\rho_\core$, to the average density of the host halo within its orbit, denoted by $\rho_\host$. These parameters determine the dimensionless angular frequency via $\rho_\core/\rho_\host = 1/(3\omega^2)$. The right panel of \cref{fig:single-field} shows the dependence of the decay rate $\tilde\Gamma$ on $\rho_\core/\rho_\host$ as computed by the BCI method, matching the fitting function presented by \refcite{Du:2018qor}. Of course, the decay rate increases with $\omega$, which fixes the tidal potential. Note that $\tilde\Gamma$ is very nearly exponential in $-\omega^{-2}$ in the regime shown. In fact, in the small-$\omega$ limit, $\Gamma$ is exponential in $-\omega^{-1}$ instead. We show this explicitly in \cref{sec:bohr-sommerfeld-appendix}, as a corollary of the results of \cref{sec:semiclassics}.

A consequence of the redefinitions we have made above is that not all of the rescaled dimensionless quantities are independent. One might think to describe a soliton with the physical quantities $\tilde m$, $\tilde\omega$, $\tilde\rho_\core$, $\tilde\rho_\host$, and $\tilde M_\soliton$.
But each of these apart from $\tilde m$ scales with a power of the scaling factor $s$, which itself depends on $\tilde m$. Since $\omega$ is the only free parameter in the rescaled system of \cref{eq:tidal-system}, only three parameters, such as $\tilde m$, $\tilde\rho_\core$, and $\tilde\rho_\host$, are needed to fix all observables. It is also possible to, e.g., replace $\tilde\rho_\core$ by $\tilde M_\soliton$. However, this is often numerically inconvenient. The only input to the numerical solution of the profile is $\omega$, and the relationship between $M_\soliton$ and $\omega$ depends on the profile itself. Thus, in practice, to find a solution corresponding to a fixed value of $M_\soliton$, it is necessary to scan over values of $\omega$. (See \refcite{Dave:2023egr} for a discussion of such a calculation.)

\section{Expectations and challenges for multifield halos}
\label{sec:two-field}
We now consider the extension of the framework of \cref{sec:single-field} to the case of a multifield halo. First, we write the equivalent of the system of \cref{eq:tidal-system} for a two-field halo, and introduce a parametrization for the boundary conditions.

\subsection{Two-field system}
\label{sec:sub:two-field-setup}
Consider a halo composed of two ultralight DM species, $\tilde\Phi_a$ and $\tilde\Phi_b$, with masses $ \tilde{m}_a$ and $ \tilde{m}_b$ respectively, coupled only gravitationally. We choose $ \tilde{m}_a < \tilde{m}_b$ in the following. We can again write $\tilde\Phi_i$ as the product of slowly- and rapidly-varying components, using $\tilde{\phi}_i$ to denote the slowly varying component. In the weak-gravity limit, we now have one equation for each field, and both fields contribute to the gravitational potential $\tilde V$ through
\begin{equation}
    \tilde\nabla^2 \tilde V=4\pi G\left(
        \bigl|\tilde\phi_a\bigr|^2 + \bigl|\tilde\phi_b\bigr|^2
    \right),
    \qquad
    \begin{cases}
        \displaystyle
        i\hbar \frac{\partial\tilde\phi_a}{\partial \tilde t}=
        -\frac{\hbar^2}{2 \tilde{m}_a}\tilde{\nabla}^2\tilde\phi_a
        + \tilde{m}_a \tilde V_{\text{eff},a}\tilde\phi_a
        ,
        \\[3mm]
        \displaystyle
        i\hbar \frac{\partial \tilde\phi_b}{\partial \tilde t}=
        -\frac{\hbar^2}{2 \tilde{m}_b}\tilde\nabla^2\tilde{\phi}_b
        + \tilde{m}_b \tilde{V}_{\text{eff},b}\tilde{\phi}_b
        .
    \end{cases}
\end{equation}
Here, as before, we take $\tilde V_{\text{eff},i}=\tilde V+\tilde V_{\text{tidal},i}$ for the subhalo potential $\tilde V$ and tidal potential from the halo $\tilde V_{\text{tidal},i}=-\frac{3}{2}\tilde\omega^2\tilde r^2$.  Again assuming a spherically symmetric, quasistationary solution of the form $\tilde\phi_i(\tilde t, \tilde r) = \tilde\chi_i(\tilde r) e^{-i\tilde\gamma_i \tilde t/\hbar}$ for each species, we obtain the system
\begin{equation}
    \tilde{\nabla}^2 \tilde{V}=4\pi G\left(
            |\tilde{\chi}_a|^2+|\tilde{\chi}_b|^2\right)
    ,
    \qquad
    \begin{cases}
        \displaystyle
        \frac{\hbar^2}{2 \tilde{m}_a}\tilde{\nabla}^2\tilde{\chi}_a = \left(
             \tilde{m}_a\tilde{V}-\tilde{\gamma}_a
            - \tfrac{3}{2} \tilde{m}_a\tilde{\omega}^2\tilde{r}^2
        \right) \tilde{\chi}_a
        ,
        \\[3mm]
        \displaystyle
        \frac{\hbar^2}{2 \tilde{m}_b}\tilde{\nabla}^2\tilde{\chi}_b = \left(
             \tilde{m}_b\tilde{V}-\tilde{\gamma}_b
            - \tfrac{3}{2} \tilde{m}_b\tilde{\omega}^2\tilde{r}^2
        \right) \tilde{\chi}_b
        .
    \end{cases}
\end{equation}
We again switch to dimensionless units. The nondimensionalization as prescribed in \cref{eq:dimensionless-variables} depends on the particle mass $\tilde m$, while we now consider a system with two mass scales, $ \tilde{m}_a$ and $ \tilde{m}_b$. We choose to nondimensionalize via $ \tilde{m}_a$, setting
\begin{equation}
    \hat{\chi}_i=\frac{\hbar\sqrt{4\pi G}}{ \tilde{m}_a c^2}\tilde{\chi}_i,
    \quad
    \hat{V}=\frac{\tilde{V}}{c^2},
    \quad
    \hat{\gamma}_i=\frac{\tilde{\gamma}_i}{ \tilde{m}_a c^2},
    \quad
    \hat{\omega}=\frac{\hbar}{ \tilde{m}_a c^2}\tilde{\omega},
    \quad
    \hat{r}=\frac{ \tilde{m}_ac}{\hbar}\tilde{r},
    \quad
    \hat{t}=\frac{ \tilde{m}_a c^2}{\hbar}\tilde{t}.
\end{equation}
Our choice of $ \tilde{m}_a$ over $ \tilde{m}_b$ is arbitrary, but influences our later definitions. In addition, this choice implies that the scaling factor of \cref{eq:scaling-factor} applies with $\tilde m_a$ in place of $\tilde m$. In terms of the dimensionless variables, we now obtain 
\begin{equation}
    \hat{\nabla}^2\hat{V}= |\hat{\chi_a}|^2+|\hat{\chi_b}|^2
    ,
    \qquad
    \begin{cases}
        \displaystyle
        \hat{\nabla}^2 \hat{\chi}_a = 2\left(
            \hat{V}- \hat{\gamma}_a - \tfrac{3}{2}\hat{\omega}^2\hat{r}^2
        \right) \hat{\chi}_a
        ,
        \\[3mm]
        \displaystyle
        \hat{\nabla}^2 \hat{\chi}_b = 2\theta\left(
            \theta\hat{V} - \hat{\gamma}_b
            - \tfrac{3}{2}\theta\hat{\omega}^2\hat{r}^2
        \right)\hat{\chi}_b
        .
        \\[3mm]
    \end{cases}
\end{equation}
where $\theta\equiv\tilde{m}_b/ \tilde{m}_a$. We will write $m_i \equiv \tilde m_i / \tilde m_a$ for the nondimensionalized form of either mass, so that $m_a = 1$ and $m_b = \theta$. Without loss of generality, we will take $\theta > 1$ unless otherwise indicated. Note that while the particle mass completely cancels in the nondimensionalized single-field system, the mass ratio $\theta$ persists in the two-field system.

The boundary conditions of the two-field system are the same as in the single-field case, for each of $\hat\chi_a$ and $\hat\chi_b$, apart from the inner boundary conditions $\hat\chi_a(0)$ and $\hat\chi_b(0)$. The central density is now given by $\hat\rho_\core=|\hat\chi_a(0)|^2+|\hat\chi_b(0)|^2\equiv \hat\rho_{\core,a}+\hat\rho_{\core,b}$, and the two-field analogue of the scaling symmetry is
\begin{equation}
    \hat t \to s^2\hat t,
    \quad
    \hat r \to s\hat r,
    \quad
    \hat\chi_i \to s^{-2}\hat\chi_i,
    \quad
    \hat V \to s^{-2}\hat V,
    \quad
    \hat\gamma_i \to s^{-2}\hat\gamma_i,
    \quad
    \hat\omega \to s^{-2}\hat\omega
   .
\end{equation}
We choose $s$ to rescale our system such that the rescaled variables satisfy $|\chi_a(0)|^2 + |\chi_b(0)|^2 = 1$, and we parametrize the relative magnitude of the two fields at the origin as
\begin{equation}
    \chi_a(0)=\frac{1}{\sqrt{1+\eta^2}},
    \qquad
    \chi_b(0)=\frac{\eta}{\sqrt{1+\eta^2}}
    ,
\end{equation}
where the \textit{central field ratio} $\eta$ is defined as $\chi_b(0)/\chi_a(0)$, or, equivalently, $\sqrt{\tilde\rho_{\core,b}/\tilde\rho_{\core,a}}$. The dimensionless, rescaled system is
\begin{equation}
    \label{eq:two-field-sys}
    \nabla^2V= |\chi_a|^2+|\chi_b|^2
    ,
    \qquad
    \begin{cases}
        \displaystyle
        \nabla^2 \chi_a = 2\left(
            V- \gamma_a - \tfrac{3}{2}\omega^2r^2
        \right) \chi_a
        ,
        \\[3mm]
        \displaystyle
        \nabla^2 \chi_b = 2\theta\left(
            \theta V - \gamma_b
            - \tfrac{3}{2}\theta\omega^2r^2
        \right)\chi_b
        ,
    \end{cases}
\end{equation}
with boundary conditions given by
\begin{equation}
    \label{eq:two-field-bc}
    \arraycolsep=5mm
    \begin{array}{llll}
        \displaystyle
        \chi_a(0)=\frac{1}{\sqrt{1+\eta^2}},
        &
        \chi_a'(0)=0,
        &
        \chi_a(\infty)=0,
        &
        V(0)=0,
        \\[5mm]
        \displaystyle
        \chi_b(0)=\frac{\eta}{\sqrt{1+\eta^2}},
        &
        \chi_b'(0)=0,
        &
        \chi_b(\infty)=0,
        &
        V'(0)=0.
    \end{array}
\end{equation}
As before, once we have obtained $\chi_i(r)$ and $\gamma_i$, the nondimensionalized mass contained within a radius $r_*$ is given by
\begin{equation}
    M_i(t,r_*) = \int_0^{r_*}\du r\,|\Phi_i(t,r)|^2
        = e^{2\Im{(\gamma_i)}t}\int_0^{r_*}\du r\,|\chi_i(r)|^2
    ,
\end{equation}
so we define the decay rate of species $i$ as
\begin{equation}
    \Gamma_i \equiv -\frac{1}{M_i}\frac{\du M_i}{\du t} = -2\Im{(\gamma_i)}.
\end{equation}
Note that each species has its own decay rate in this definition, meaning that the decay rate of the total mass of the halo is time-dependent: once the faster-decaying component of the soliton has mostly evaporated away, the total decay rate becomes that of the longer-lived component. We neglect this time dependence. Realistically, over long timescales, even a single-field halo has a time-dependent decay rate: mass loss modifies $\rho_\core$, which modifies $\omega$. Accordingly, the time dependence in the overall decay rate cannot be extracted from a simple comparison of $\Gamma_a$ and $\Gamma_b$. In the rest of our analysis, we treat the decay rates as time-independent, order-of-magnitude indicators of cosmological stability or instability, but refer to \refcite{Du:2018qor} for a numerical study of time dependence in the single-field case.

Having now defined the appropriate variables for a two-field halo, we consider how we expect $\gamma_i$ (and thus $\Gamma_i$) to behave as functions of $\omega$, $\theta$, and $\eta$. Direct numerical solution of the two-field system is quite challenging, but several limiting cases are immediately tractable.

\subsection{Limiting cases}
\label{sec:sub:two-field-predictions}
First, consider the case of large $\theta$ at fixed $\eta>1$. The length scale of the soliton radius for each field is set by its Compton wavelength, $\tilde\lambda_i \propto \tilde m_i^{-1}$. In the limit of large $\theta$, the heavier field is effectively confined to the center, since $\lambda_b/\lambda_a \to 0$. Here, at radii much smaller than the length scale of variation of the lighter field, $\chi_a$ can be treated as a uniform background with its central value $(1+\eta^2)^{-1/2}$. The equations of motion for $\chi_b$ then reduce to
\begin{equation}
    \nabla^2 \chi_b = 2\theta\left(
        \theta V - \gamma_b - \theta\tfrac{3}{2}\omega^2r^2
    \right) \chi_b
    ,
    \qquad
    \nabla^2V \approx |\chi_b|^2+\frac{1}{1+\eta^2}
    .
\end{equation}
Now, observe that we can relate this to the single-field scenario. Let $V_i$ denote the potential sourced by field $i$, so that $V=V_a+V_b$ and $\nabla^2V_b = |\chi_b|^2$. Then we have $V_a = r^2/[6(1+\eta^2)]$. The equation of motion for $\chi_b$ then takes the form
\begin{equation}
    \nabla^2 \chi_b = 2\theta\left[
        \theta V_b - \gamma_b - \theta\left(\frac{3}{2}\omega^2+\frac{1}{6\,\theta(1+\eta^2)}\right)r^2
    \right] \chi_b
    .
\end{equation}
It follows that the solution for the profile of the heavy field corresponds to a \textit{single-field} solution with a modified angular frequency satisfying
\begin{equation}
    \omega'^2 = \omega^2 + \frac{1}{9\,\theta(1+\eta^2)}
    .
\end{equation}

We can also qualitatively predict the behavior of the light field in the same limit by examining the potential $V_b$ generated by the heavy field, since $V_b$ parametrizes the difference between the two-field system and a one-field system with only the light field $\chi_a$. Since we have established that $\chi_b$ is well approximated by a single-field soliton profile in this regime, the potential $V_b$ can be estimated from the fitting formula for the single-field soliton profile, \cref{eq:single-field-profile}. Rescaling this approximation, we have $\chi_b(r) \approx \eta(1+\eta^2)^{-1/2} \exp(-ar^2\theta^2)$ for $a=0.1815$, which gives
\begin{equation}
    V_b \approx \frac{\eta^2}{1+\eta^2}\left[
        \frac{4r\theta\sqrt{a}
            - \sqrt{2\pi}\erf\left(r\theta\sqrt{2a}\right)}
        {16r\left(\theta\sqrt a\right)^3}
    \right]
    .
\end{equation}
In the large-$r$ limit, $V_b$ has the form $V_\infty - A/r$ for constants
\begin{equation}
    \label{eq:long-range-potential-parameters}
    V_\infty = \frac{\eta^2}{1+\eta^2}\frac{1}{4a\theta^2}
    \text{~~and~~}
    A = \sqrt{\frac\pi2}\frac{\eta^2}{1+\eta^2}\frac{1}{8a^{3/2}\theta^3}
    ,
\end{equation}
which reflects the fact that the heavy field contributes to the potential like a point mass at long range, with effective mass proportional to $A$. In turn, $A\propto\theta^{-3}$ because the mass is parametrically the product of the central density, $\eta^2(1+\eta^2)^{-1}$, and the volume within the Compton wavelength, $\lambda_b\sim\theta^{-1}$. Now, $V_b$ exhibits two important limits at fixed $\theta$:
\begin{equation}
    V_b \approx \frac{\eta^2}{1+\eta^2}
    \begin{cases}
        \tfrac16r^2 & r \ll \theta^{-1}a^{-1/2},
        \\
        \tfrac14a^{-1}\theta^{-2} & r \gg \theta^{-1}a^{-1/2}.
    \end{cases}
\end{equation}
This is the contribution of the heavy field to the potential that governs the dynamics of the light field. In particular, for small $r$, the light field is subject to a \textit{reduced} tidal potential, with an effective angular frequency given by
\begin{equation}
    \omega'^2 = \omega^2 - \frac{\eta^2}{9(1+\eta^2)}
    .
\end{equation}
This reflects the fact that the denser core binds the soliton more tightly. In practice, however, the tidal stripping rate is dominated by the behavior at larger $r$, where the effect of the heavy field reduces to a shift in the eigenvalue $\gamma_a$.

\begin{figure}
    \centering
    \includegraphics[width=\linewidth]{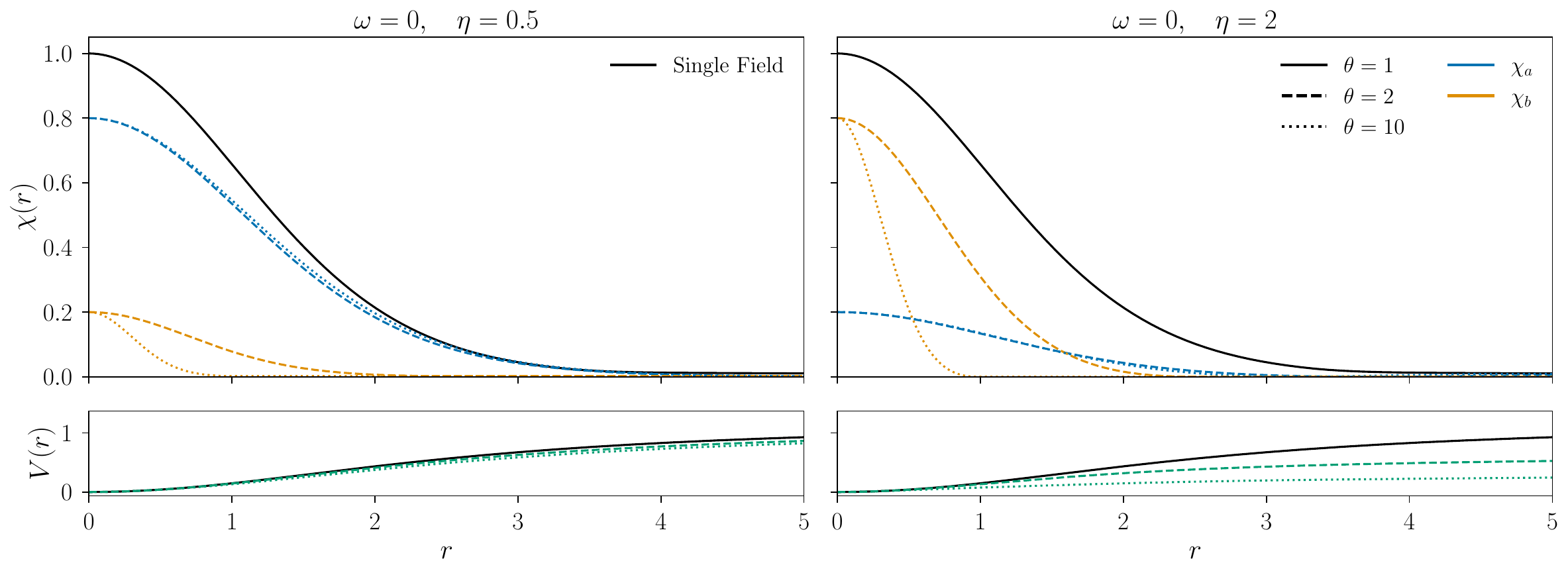}
    \caption{Profiles and potentials for several values of the mass ratio $\theta$, in the absence of a tidal potential ($\omega = 0$). The black curve shows the single-field solution. The two panels differ only by the central field ratio $\eta$, which is taken to be $0.5$ and $2.0$ in the left and right panels, respectively. In the right panel, where both $\theta>1$ and $\eta>1$, each field dominates the density over some range of radii.}
    \label{fig:subdominance}
\end{figure}

Several limits now follow directly from this analysis, which we list below. For illustrative purposes, \cref{fig:subdominance} shows the profiles of several relevant cases, obtained using the solution process to be described in \cref{sec:numerics}. The key limits are as follows:
\begin{enumerate}
    \item $\theta\gg1$, $\eta\gg1$ (dotted curves in the right panel of \cref{fig:subdominance}): The profile of $\chi_b$ matches an ordinary single-field solution throughout its Compton wavelength. This is simply because the density is dominated by $\chi_b$ over this entire region. The inner portion of the light field's profile receives a correction to $\omega^2$ of order $-1/9$, reducing the effective tidal potential. The outer portion receives a correction to the eigenvalue relative to the single-field case.
    \item $\theta\gg1$, $\eta\ll1$ (dotted curves in the left panel of \cref{fig:subdominance}): The density of $\chi_a$ always dominates over $\chi_b$. The profile of $\chi_a$ matches an ordinary single-field solution, whereas $\chi_b$ receives a contribution to $\omega^2$ of order $(9\theta)^{-1}$. Since we are often interested in $\omega^2\sim\num{e-3}$ and $\theta\sim10$, such a contribution can be significant even when $\theta$ is large.
    \item $\theta=1$ (black curves in both panels of \cref{fig:subdominance}): The equations of motion for the two fields are identical, with rescaled boundary conditions. This scenario is equivalent to the single-field scenario.
    \item $\theta<1$: The equations of motion are symmetric under the interchange of the two fields accompanied by the transformation $(\theta, \eta) \to (1/\theta,1/\eta)$. Thus, in this case, fields $a$ and $b$ with $\eta\gg1$ and $\eta\ll1$ exhibit the same behavior as fields $b$ and $a$ in the $\theta\gg1$ case described above with $\eta\ll1$ and $\eta\gg1$, respectively.
\end{enumerate}

\section{Numerical methods for real eigenvalues}
\label{sec:numerics}
Our primary goal in this work is to extract tidal stripping rates, which correspond to the imaginary part of the eigenvalue $\gamma$. The standard method to solve for $\Im(\gamma)$ is to solve the full quasistationary eigenvalue problem with the BCI method, and simply extract the imaginary part. This approach is well established in the single-field case, both for $\omega=0$ (i.e., without tidal stripping)~\cite{Marsh:2015wka} and $\omega > 0$~\cite{Hertzberg:2022vhk, Dave:2023egr}. It has even been used previously in the multifield case for $\omega=0$~\cite{Luu:2018afg, Luu:2023dmi}. However, this approach suffers from severe numerical difficulties~\cite{vanDissel:2023vhu}. It is particularly difficult to solve the system for values of $\theta$ that are not close to 1, even for $\omega=0$, as noted by \refcite{Luu:2023dmi}. Here we introduce methods to robustly determine only the \textit{real} part of the eigenvalue. We will show in \cref{sec:semiclassics} that the imaginary part of $\gamma$ can be determined directly from the real part, meaning that the full complex eigenvalue can be determined for $\omega > 0$.

\subsection{Boundary condition at infinity}
First, we show that it is possible to easily match boundary conditions at infinity when $\gamma$ is purely real. We begin by explaining why the numerical problem is difficult, and especially so for the imaginary part. The system of \cref{eq:tidal-system} represents a nonlinear problem with an undetermined eigenvalue $\gamma$. Here, the simplest numerical approach is the shooting method: a residual loss $L(\gamma)$ is defined in terms of the deviation from the boundary conditions, such that $L(\gamma) = 0$ for a valid eigenvalue. Then solving for $\gamma$ is a nonlinear root-finding problem. When $\omega = 0$, so that $\gamma$ is real, this is a relatively straightforward one-dimensional problem. However, the presence of a tidal potential means that the eigenvalue $\gamma$ is complex, leading to a much more difficult two-dimensional root-finding problem. Two fields in a tidal potential give rise to a four-dimensional root-finding problem, which is often intractable in itself.

Worse yet, the numerical problem is generically ill-conditioned in the regime of interest. In order to correctly define the boundary conditions at infinity, a sufficiently large maximum radius must be chosen. This is particularly important for the case where $\omega > 0$, since matching to the WKB solution at large radii requires a substantial segment in the regime of validity of the large-$r$ solution. But at large radii, a small deviation from the precise value of $\gamma$ produces exponential growth where the solution must instead decay as a power law. As a result, the residual of the boundary condition at infinity is extraordinarily sensitive to the \emph{real} part of $\gamma$. For example, in our nondimensionalized units, with a maximum radius of 20, a deviation in $\gamma$ at the level of one part in \num{e8} is enough to cause substantial disagreement with the boundary conditions at the maximum radius, and a deviation of one part in \num{e5} is enough to cause exponential growth so severe that numerical integration cannot be successfully completed with ordinary-precision floating point arithmetic. This extreme sensitivity to the real part almost completely masks all sensitivity to the imaginary part, making it very difficult to determine the quantity $\Im(\gamma)$ that actually controls the rate of tidal stripping.

It is easy to understand why the solution is so much more sensitive to $E \equiv \Re(\gamma)$ than to $\Gamma \equiv -2\Im(\gamma)$. In terms of these variables, \cref{eq:tidal-system-re-im} takes the form
\begin{equation}
    \begin{array}{l}
        (2/r)\partial_r\Re(\chi) + \partial_r^2\Re(\chi) =
            2\left[V(r) - E
                - \tfrac32\omega^2r^2\right]\Re(\chi)
            - \Gamma\Im(\chi)
        ,
        \\[2mm]
        (2/r)\partial_r\Im(\chi) + \partial_r^2\Im(\chi) =
            2\left[V(r) - E
                - \tfrac32\omega^2r^2\right]\Im(\chi)
            + \Gamma\Re(\chi)
        ,
    \end{array}
\end{equation}
and the only role of $\Gamma$ is to couple $\Re(\chi)$ and $\Im(\chi)$. The boundary conditions provide that $\chi(0)$ is entirely real, and $\Re(\chi)$ dominates over $\Im(\chi)$ for most of the radii that contribute significantly to the mass (and therefore the potential). But this means that $\Re(\chi)$---and so, to good approximation, $V(r)$---is only influenced by $\Gamma$ at second order, by backreaction of $\Im(\chi)$ on $\Re(\chi)$ via another factor of $\Gamma$. In all scenarios of cosmological interest, we have $|\Gamma| \ll 1$, so the effect of $\Gamma$ on the structure of the profile and potential is minimal.

However, the fact that the determination of $\Gamma$ is so sensitive to the boundary condition at infinity is only an idiosyncracy of the shooting method. Considered as a tunneling problem, the decay rate is only physically sensitive to the behavior of the field and the potential near the classical turning points, which lie at much smaller radii. Moreover, the relative \emph{insensitivity} of the boundary condition to $\Gamma$ points to an important and generic result: the determination of the real part $E$ is insensitive to $\Gamma$. That is, in the minimization of $|r(\gamma)|$, if $\Gamma$ is constrained to any particular value, the value of $E$ resulting from the minimization is almost unchanged. In particular,
\begin{equation}
    \argmin_{\gamma\in\mathbb R}|r(\gamma)|
    \approx \Re\{\argmin_{\gamma\in\mathbb C} |r(\gamma)|\}.
\end{equation}

We have verified this explicitly for a single-field soliton, finding that the values of $E$ extracted from quasistationary BCI solutions (i.e., with $E = \Re\{\argmin_{\gamma\in\mathbb C}|r(\gamma)|\}$) very nearly agree with those extracted from approximate \emph{stationary} BCI solutions (i.e., with $E \approx \argmin_{\gamma\in\mathbb R}|r(\gamma)|$). Thus, as long as the \emph{real part} of $\gamma$ can be determined under the assumption that $\Im(\gamma) = 0$, the structure of the profile and the potential barrier can be reliably recovered. We will show in \cref{sec:semiclassics} that these results can subsequently be used to identify the tunneling rate, without solving directly for the imaginary part of $\gamma$. The only challenge, then, is to reliably compute the real parts of both eigenvalues for a two-field soliton.

\begin{figure}\centering
    \includegraphics[width=\linewidth]{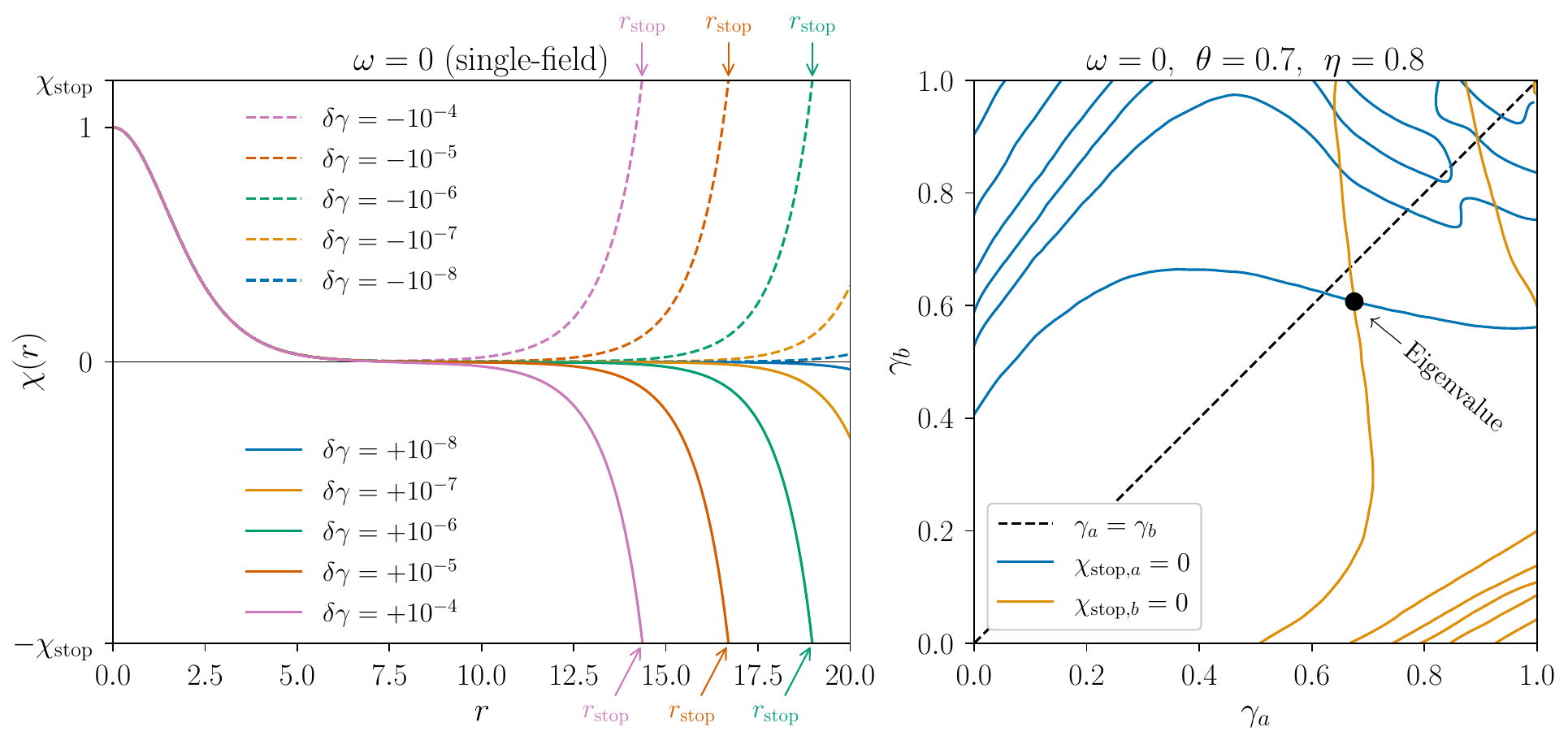}
    \caption{\ignorespaces
        Overview of the numerical procedure (see text).
        \textit{Left:} illustration of the determination of $r_{\mathrm{stop}}$, shown here for a single-field system with $\omega = 0$. Each curve shows $\chi(r)$ computed with a different value for the deviation $\delta\gamma$ from the true value of $\gamma$, which is purely real for $\omega = 0$ (i.e., $\gamma = E$). Note the severe (exponential) sensitivity to $r\times\delta\gamma$.
        \textit{Right:} identification of eigenvalue pair from intersecting contours, shown here for a two-field halo with $\theta = 0.7$ and $\eta = 0.8$. Since $\theta$ and $\eta$ differ slightly from 1, the eigenvalues are slightly unequal.
    }
    \label{fig:numerical-procedure}
\end{figure}

Fortunately, the same behavior that makes the direct determination of the imaginary parts difficult makes the determination of the real parts numerically easy. When $E$ slightly deviates from the real part of the true eigenvalue, then $|\chi(r)|$ diverges rapidly at large $r$. Any solution with $|\chi(r)| > 1$ is unphysical, and on either side of the correct eigenvalue, $\chi(r)$ will attain a value greater than 1 or less than $-1$ for some finite $r$, generally quite small. The correct value of the real part can then be determined by a simple numerical procedure: \emph{terminate} numerical solution of the system immediately when $|\chi(r)| = \chi_{\mathrm{stop}}$ for some $\chi_{\mathrm{stop}} > 1$, and denote the corresponding value of $r$ by $r_{\mathrm{stop}}$. Then, generically, $\chi(r_{\mathrm{stop}}) = \pm\chi_{\mathrm{stop}}$, and the sign is a function of $\gamma$, with a sharp transition between positive and negative values taking place at an eigenvalue. This behavior is illustrated in \cref{fig:numerical-procedure}. Finding the eigenvalue is then a simple root-finding problem that can be numerically performed very rapidly to high precision. We refer to this as the \textit{stationary} BCI method, since the eigenvalue is restricted to be real, but the boundary condition is still enforced at large $r$.

More importantly, this procedure easily extends to the case of multiple fields. In the original eigenvalue problem, the added difficulty for multiple fields arises from the fact that the residual loss $L(\gamma)$ between the candidate solution and the boundary conditions becomes a function of multiple (complex) variables, $L(\gamma_a,\gamma_b,\dotsc)$, with great numerical sensitivity to the real part of each eigenvalue. Na\"ively solving $|L| = 0$ is thus numerically problematic. However, in slightly modified form, the above procedure can be factorized and iterated over several fields. Specifically, we implement the following numerical procedure for two fields:
\begin{enumerate}
    \item For each combination of real-valued $\gamma_a$ and $\gamma_b$, evaluate $\chi_a(r_{{\mathrm{stop}},a})$ and $\chi_b(r_{{\mathrm{stop}},b})$ as above. Scan over the relevant region of $(\gamma_a, \gamma_b)$.
    \item Identify the contours separating positive and negative values of each of $\chi_a$ and $\chi_b$.
    \item Find the intersection point of these contours with least total energy.
\end{enumerate}
This intersection point is the desired eigenvalue pair $(E_a,E_b)$. The identification of $\chi_a(r_{\mathrm{stop}})$ is illustrated in the left panel of \cref{fig:numerical-procedure}, and the intersection procedure is shown in the right panel.

There is one complication that must be addressed in the implementation of this procedure: suppose that $r_{{\mathrm{stop}},a} > r_{{\mathrm{stop}},b}$. Then, in order to evaluate $\chi(r_{{\mathrm{stop}},a})$, the system must be integrated beyond $r_{{\mathrm{stop}},b}$, where $\chi_b$ is badly divergent. Since $\chi_a$ and $\chi_b$ are coupled via the gravitational potential, this disrupts the computation of $\chi_a$. However, typically, if $ \tilde{m}_a <  \tilde{m}_b$ (i.e., $\theta > 1$), then on physical grounds, the more centrally-concentrated $\chi_b$ should fall nearly to zero before $r_{{\mathrm{stop}},a}$. Thus, we can use the following procedure:
\begin{enumerate}
    \item Integrate the coupled system to determine both $\chi_a$ and $\chi_b$ for $r < r_{{\mathrm{stop}},b}$.
    \item Identify the radius $r_m$ of the last stationary point of $\chi_b$, and \emph{truncate} beyond this radius, setting $\chi_b(r) = 0$ for $r > r_m$. If $\chi_b(r)$ has no stationary points, set $r_m$ to the last stationary point of $\chi_b'(r)$.
    \item Resume integration for $\chi_a$ alone to determine $\chi_a(r_{{\mathrm{stop}},a})$.
\end{enumerate}
This truncation procedure is justified by the fact that \emph{for the correct eigenvalue pair,} it is indeed the case that $\chi_b$ should be negligible beyond $r_m$. Thus, the pair $(E_a, E_b)$ obtained by the subsequent intersection procedure is one for which the only effect of the truncation is to remedy the large-$r$ sensitivity to very small deviations in $E_b$.

This procedure is numerically inexpensive and very robust at small $\omega$. For $\omega \gtrsim 0.5$, numerical noise becomes significant for some combinations of $\theta$ and $\eta$ due to the high required precision on $E_i$. We note, however, that the $\omega$ dependence is very well fit by a function of the form
\begin{equation}
    \label{eq:tidal-response}
    E_i = A_i\exp(-\alpha_i\omega^2)
    ,
\end{equation}
in which we call $A_i$ the \textit{isolated eigenvalue parameter,} since it corresponds to $\gamma_i$ at $\omega=0$, and we call $\alpha_i$ the \emph{tidal response parameter,} since it fully characterizes the dependence of the soliton structure on the tidal potential. Since $A_i$ is easily computed directly at $\omega=0$, the fit need only be performed in the single parameter $\alpha_i$. We will ultimately fit $\alpha_i$ for several combinations of $\theta$ and $\eta$, obtaining numerically-stable values for $E_i$ for each such parameter point across a wide range of angular frequencies $\omega$. The single-field solution (corresponding to $\theta=1$) is characterized by
\begin{equation}
    \label{eq:single-field-tidal-response}
    A = 0.6495,\qquad \alpha = 9.12
    .
\end{equation}
We show in \cref{sec:bohr-sommerfeld-appendix} that the small-$\omega$ behavior of this result can be predicted analytically, meaning that this value of $\alpha$ can be determined approximately from semiclassical methods without explicit fitting.

\subsection{Bohr-Sommerfeld quantization}
\label{sec:bohr-sommerfeld}
The procedure of the previous section makes the two-field eigenvalue problem tractable, but it still involves considerable complexity, and the truncation procedure can introduce its own numerical challenges for certain parameter points. It is therefore desirable to have a simpler alternative method of approximating the eigenvalues, ideally one that is not sensitive to the behavior of the profile at large $r$. In fact, there is a standard approximation in which it is not necessary to enforce the boundary condition at infinity at all. The real part of the eigenvalue can be determined while working strictly in the \textit{interior} of the potential well by means of Bohr-Sommerfeld quantization.

The Bohr-Sommerfeld quantization procedure translates the boundary condition at infinity to a condition on the WKB action in the well, requiring that the eigenvalue satisfies the \textit{Bohr-Sommerfeld quantization condition,}
\begin{equation}
    \label{eq:bohr-sommerfeld-rule}
    \int_0^{r_1}\du r\,\sqrt{
        2m\left[E-mV_\eff(r)\right]
    } =
    \left(n + \frac{3}{4}\right)\pi \text{~~for~} n\in\{0,1,2,\dotsc\}
    ,
\end{equation}
where $r_1$ is the inner turning point. We discuss WKB methods in more detail in \cref{sec:semiclassics}, and we review the derivation and other implications of the Bohr-Sommerfeld quantization condition in \cref{sec:bohr-sommerfeld-appendix}. For present purposes, we observe that the Bohr-Sommerfeld condition can be evaluated easily, especially for large $\omega$, since the system need not be integrated beyond the first classical turning point.

\begin{figure}
    \centering
    \includegraphics[width=\textwidth]{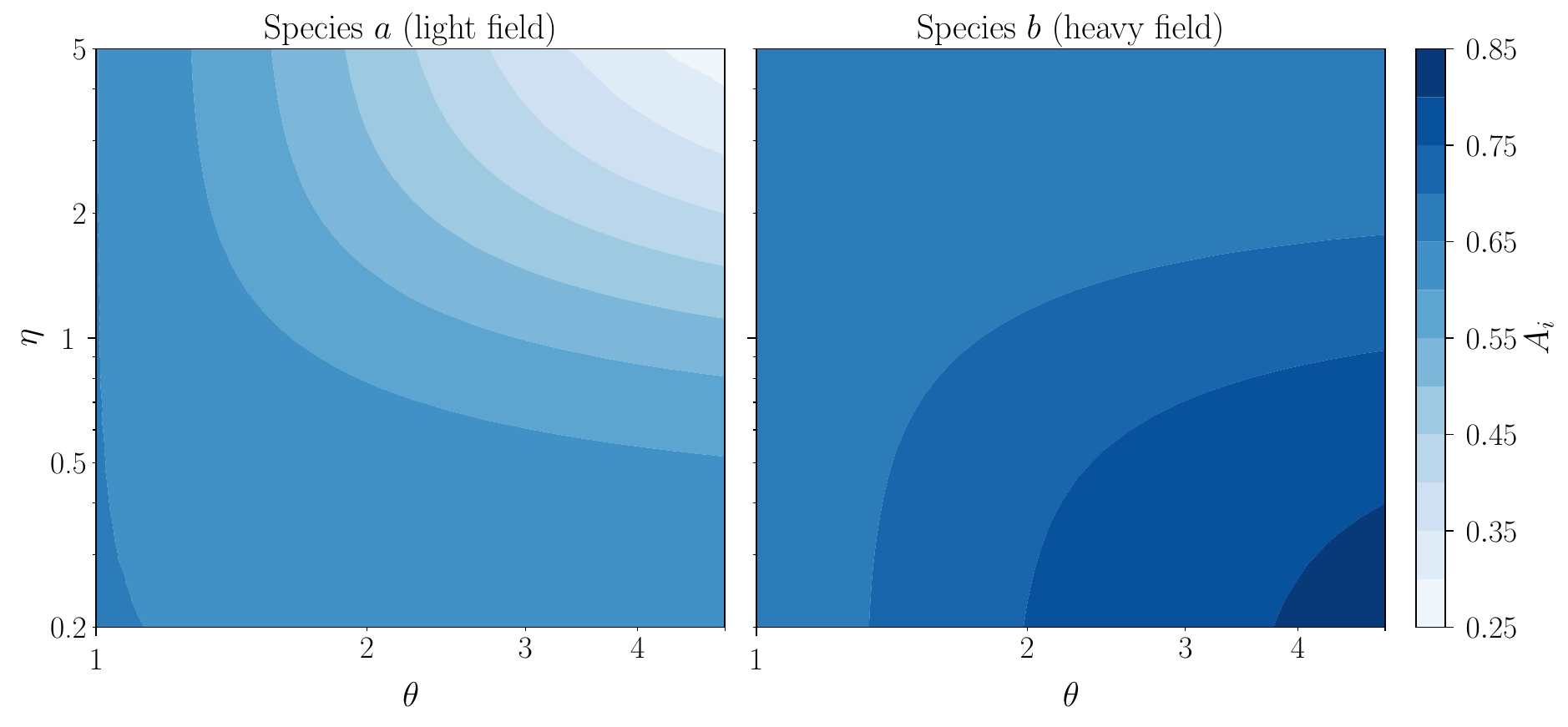}
    \caption{Isolated eigenvalue parameter $A_i$, i.e., $E_i$ at $\omega=0$, as determined by Bohr-Sommerfeld quantization for the $a$ field ($m=1$, left) and the $b$ field ($m=\theta$, right), scanning over the mass ratio $\theta$ and the central field ratio $\eta$.}
    \label{fig:re-gamma-theta-eta}
\end{figure}

\begin{figure}
    \centering
    \includegraphics[width=\textwidth]{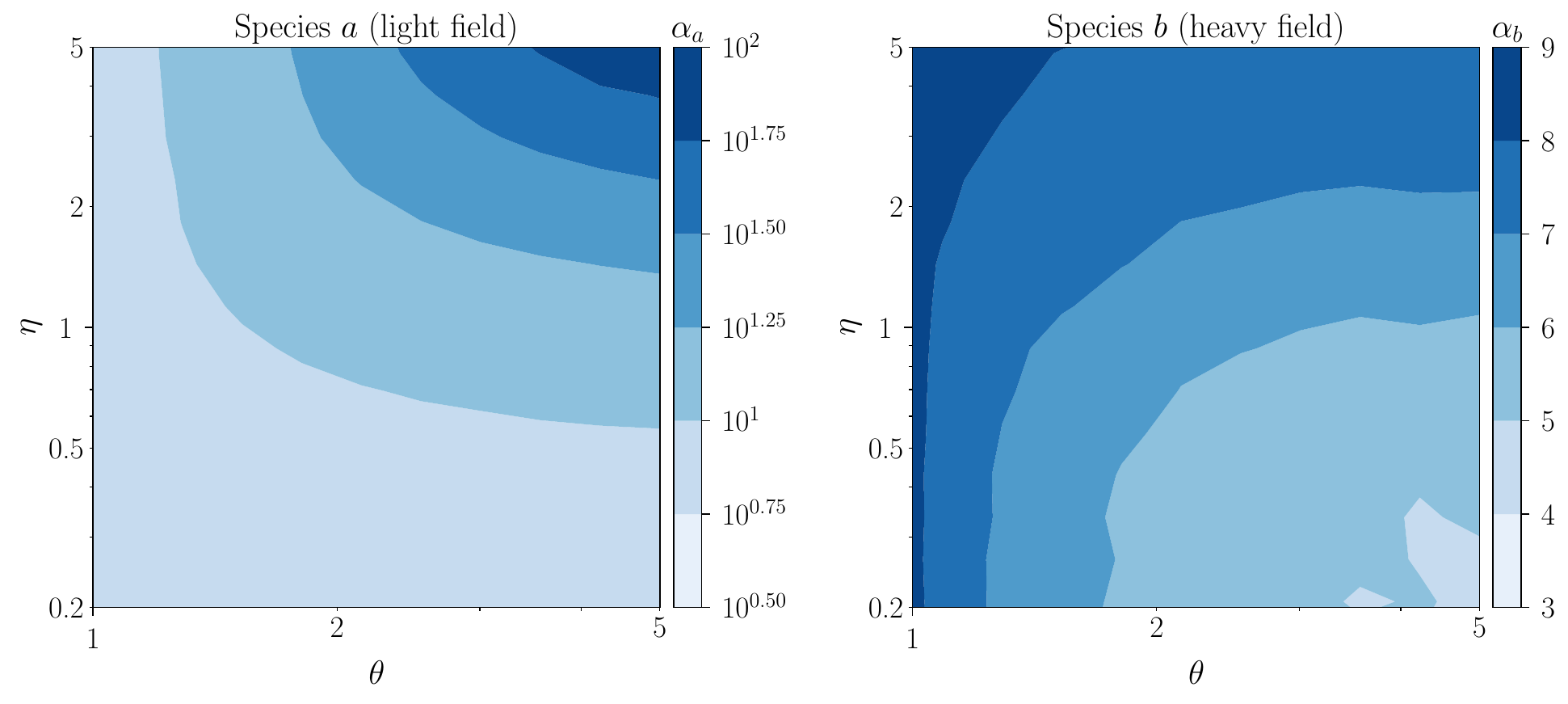}
    \caption{Tidal response parameter $\alpha_i$ as determined by Bohr-Sommerfeld quantization for the $a$ field ($m=1$, left) and the $b$ field ($m=\theta$, right), scanning over the mass ratio $\theta$ and the central field ratio $\eta$. The heavy field exhibits a narrow range of $\alpha$ values, and the corresponding plot is shown with a linear color scale.}
    \label{fig:alpha-theta-eta}
\end{figure}

In particular, this means that the eigenvalue can be approximated without ever numerically imposing a boundary condition at large $r$. Since much of the numerical difficulty inherent in solving for the eigenvalues relates to the sensitivity of the large-$r$ behavior to small changes in the eigenvalue, the Bohr-Sommerfeld approach sidesteps the most problematic obstruction entirely, making it much easier to quickly study large numbers of parameter points. \Cref{fig:re-gamma-theta-eta} shows the isolated eigenvalue parameter $A_i$ (i.e., $E_i$ at $\omega = 0$), as a function of $\theta$ and $\eta$. At $\theta = 1$, we find $A_i\approx 0.65$ for both fields regardless of $\eta$, matching \cref{eq:single-field-tidal-response}.

Similarly, \cref{fig:alpha-theta-eta} shows the tidal response parameter $\alpha_i$ as computed via the Bohr-Sommerfeld procedure, i.e., by fitting to a set of eigenvalues computed with varying $\omega$. As expected, we have $\alpha_i \approx 9$ in the limit $\theta\to 1$ for both fields, again matching \cref{eq:single-field-tidal-response}. We further have $(A_a, \alpha_a) \approx (0.65, 9)$ in the limit $\eta\to0$, and $(A_b, \alpha_b)\approx (0.65, 9)$ in the limit of large $\eta$, each regime corresponding to a single-field--dominated solution. The tidal response parameter $\alpha_b$ becomes small for the heavy field at large $\theta$ and small $\eta$, where the potential is dominated by the nearly-uniform lighter field. On the other hand, $\alpha_a$ becomes large for large $\theta$ and large $\eta$, a regime in which the heavy field dominates the potential at small radii. In this case, the self-gravity of the lighter field is only relevant at larger radii, where the tidal potential is stronger. Together, \cref{fig:re-gamma-theta-eta,fig:alpha-theta-eta} give a full description of the behavior of $E_i \equiv \Re(\gamma_i)$ in two-field halos under the Bohr-Sommerfeld approximation.

Having now developed methods to compute the real parts of the eigenvalues for both fields, we turn to the imaginary parts, and show that they can be computed post-hoc.

\section{Tidal stripping rates from semiclassical methods}
\label{sec:semiclassics}
In ultralight DM halos, where the Gross--Pitaevskii--Poisson equation applies, tidal stripping is analogous to quantum tunneling of dark matter particles out of the subhalo. This suggests that stripping can be efficiently described using techniques developed for the purpose of understanding tunneling more generally. Here, single-particle tunneling can be used as a mathematical analogy: while we do not describe the tunneling of individual DM particles, the dynamics of the soliton profile as a whole can be mapped directly to the wave function in simple tunneling problems. In particular, for the small tunneling rates expected of cosmologically-stable halos, a particularly efficient treatment is possible via \textit{semiclassical methods,} including the WKB approximation and instanton techniques. To our knowledge, these methods have been applied to the tidal stripping of solitons in any form only once, in an appendix of \refcite{Dave:2023egr}, which makes a concise order-of-magnitude comparison between the WKB method and the results of the quasistationary BCI computation.

Semiclassical methods generally invoke the lowest orders of an expansion in $\hbar$, and are valid when quantum effects are subdominant. In this section, we will review two semiclassical methods and demonstrate their equivalence. We will ultimately arrive at two convenient, equivalent relations for the mass loss rate $\Gamma$ as defined in \cref{eq:Gamma}, in the form 
\begin{equation}
    \frac{1}{\tilde M}\frac{\du \tilde M}{\du \tilde t}
        = \tilde\Gamma \simeq
            \tilde{\nu} \cdot \exp \left(
                - \frac{2}{\hbar} \int_{\tilde r_1}^{\tilde r_2} \du \tilde{r}
                \sqrt{2\tilde{m} (
                  \tilde{m}\tilde{V}_\eff(\tilde r)-\tilde{E}} )
                \right)
        =  \tilde{\nu} \cdot \exp \left(
            - \frac{2}{\hbar} \tilde{S}_\euc[\bar{r}_I] \right)
    ,
\end{equation}
where we use dimensionful variables here to make the $\hbar$ dependence manifest. Here $\tilde{S}_\euc$ is the Euclidean action; $\tilde r_1$ and $\tilde r_2$ are the classical turning points; $\bar{r}_I$ is a solution of the Euclidean equations of motion in the potential $\tilde{V}_\eff$ connecting $\tilde r_1$ to $\tilde r_2$;\footnote{Note that it is also common in the literature to express the tunneling rate through the Euclidean action of the \textit{bounce} solution $\bar{r}_b$, i.e., the instanton describing a motion from $r_1$ to $r_2$ and back. This accounts for a factor of $2$ in the exponent, i.e., $\tilde{S}_\euc[\bar{r}_b]=2 \tilde{S}_\euc[\bar{r}_I]$.} and the \textit{attempt frequency} $\tilde\nu$ is given by
\begin{equation}
    \label{eq:nu-bar}
	\tilde{\nu} = \frac{1}{4\tilde m} \left[
        \int_{0}^{\tilde r_1}
            \frac{\du\tilde r}{\sqrt{2\tilde m\bigl|
                \tilde m\tilde V_\eff(\tilde r)-\tilde{E}\bigr|}}
    \right]^{-1}
    .
\end{equation}
The attempt frequency is so named because it can be identified with the rate at which the semiclassical point particle trapped in the false vacuum basin collides with the barrier~\cite{Du:2018qor, Dave:2023egr, griffiths_introduction_2018, Shankar:102017}. Note that our conventions deviate from the ones commonly used in the context of more formal studies of quantum tunneling, e.g. by the factor of $\tilde{m}$ multiplying the potential. Additionally, since the energy $\tilde E$ is a parameter of the entire soliton rather than of a single particle, it is typically on the same order as $\tilde V_\eff$, and cannot be neglected in \cref{eq:nu-bar}.

We will find that relying on this method allows us to avoid one of the biggest challenges linked to BCI methods: their reliance on boundary conditions at spatial infinity, whose implementation can prove numerically challenging. The semiclassical methods make it manifest that $\tilde{\Gamma}$ is predominantly controlled by the form of the potential barrier between the two turning points $\tilde r_1$ and $\tilde r_2$, which must be the case physically.

\subsection{The WKB approximation}

The WKB approximation describes a technique for approximately solving the time-independent Schr\"{o}dinger equation for a given potential. For a point particle in one spatial dimension, this equation takes the simple form
\begin{equation}
    \label{eq:time-independent-schrodinger-1d}
    \left(-\frac{\hbar^2}{2\tilde m}\tilde \partial_{\tilde x}^2
        + \tilde m \tilde V(\tilde x) \right) \tilde \psi(\tilde x)
    = \tilde{\gamma}\tilde\psi(\tilde x)
    .
\end{equation}
For an exact eigenstate, $\gamma=E$ is strictly real. In the context of quantum tunneling, meanwhile, it is common to study so-called \textit{resonance states}---i.e., approximate eigenstates localized within a single well. For such states, $\gamma$ is generically complex, with the imaginary part of the energy being related to the tunneling rate we wish to calculate, as $\gamma=E-i\Gamma/2$. We will find, however, that in the semiclassical limit, this imaginary part is exponentially suppressed relative to $E$.

The WKB approximation sets out from a simple ansatz for the wave function,
\begin{equation}
    \psi(x) = A(x) e^{i\varphi(x)}
    .
\end{equation}
The functions $\varphi$ and $A(x)$ can then be obtained analytically by solving \cref{eq:time-independent-schrodinger-1d} to leading and next-to-leading order in $\hbar$. In terms of the auxiliary function $p(x)\equiv \sqrt{2m[E-mV(x)]}$, they can be found as~\cite{Wentzel:1926aor,Kramers:1926njj,Brillouin:1926blg}
\begin{equation}
    \label{eq:phi-wkb}
    \tilde \varphi(\tilde x) =
        \pm\frac 1\hbar\int^{\tilde{x}}\du
            \tilde x'\ \tilde p(\tilde x^\prime)
        + \mathcal{O}(\hbar),
    \qquad
    \tilde A(\tilde x) = \mathcal{N}\cdot \tilde p^{-1/2}(\tilde x),
\end{equation}
with some normalization factor $\mathcal{N}$. (The use of $E$ in place of $\gamma$ relies on the assumption that $|\Gamma| \ll |E|$, which we will justify shortly.) \Cref{eq:phi-wkb} makes immediately evident that this leading-order term is dominant in regions in which $\bigl|\tilde p(\tilde x)\bigr|\gg \hbar$, i.e., away from the classical turning points where $mV(x)=E$.

Away from these points, \cref{eq:phi-wkb} gives rise to two qualitatively different kinds of behavior for the wave function. In the classically-allowed region where $mV(x)<E$, $\varphi(x)$ is real, thus giving rise to an oscillatory wave function. In regions where $mV(x)>E$, $\varphi(x)$ is imaginary, corresponding to a superposition of an exponentially growing and decaying wave function. Specifically, consider a system with a potential barrier such that $E < mV(x)$ for $x_1 < x < x_2$ (i.e., with classical turning points $x_1$ and $x_2$). Assuming the wave function to be initially localized to the left of the barrier at $x<x_1$ and the semiclassical expansion to be valid, the WKB solution can be written as
\begin{equation}
    \tilde \psi(\tilde x)\approx
    \frac{1}{\sqrt{\tilde p(\tilde x)}}
    \begin{cases}
        \displaystyle
        C_1 \exp\left[\frac{i}{\hbar}\int\du\tilde x\,
            \tilde p(\tilde x)\right]
            + C_2 \exp\left[-\frac{i}{\hbar}\int\du\tilde x\,
            \tilde p(\tilde x)\right] & 0<\tilde x<\tilde x_1,
        \\[4mm]
        \displaystyle
        C_3 \exp\left[-\frac{1}{\hbar}\int\du\tilde x\,
            |\tilde p(\tilde x)|\right]
            & \tilde x_1<\tilde x<\tilde x_2,
        \\[4mm]
        \displaystyle
        C_4 \exp\left[\frac{i}{\hbar}\int\du\tilde x\,
            \tilde p(\tilde x)\right]
            & \tilde x_2<\tilde x.
    \end{cases}
    \label{eq:wkb-solution}
\end{equation}
For our purposes, the precise shape of the wave function near the turning points where the WKB approximation breaks down is unimportant. See \refcite{Berry:1972na} for further details.

While directly obtaining the tunneling rate $\Gamma$ from the shape of the wave function is nontrivial~\cite{Andreassen:2016cvx}, a reliable leading-order estimate can be obtained by multiplying the \textit{transmission amplitude} $\mathcal{A}\equiv|C_4|^2/|C_1|^2$ by the rate $\nu$ at which a classical particle with energy $E$ (and thus momentum $p(E)$) would hit the barrier. Using \cref{eq:wkb-solution}, it is now straightforward to find
\begin{equation}
    \label{eq:width-estimate}
    \tilde \Gamma \sim \frac{p(E)}{2m} \mathcal{A} \simeq
    \tilde{\nu} \exp \left(
        - \frac{2\tilde S_\euc}{\hbar}\right),
    \qquad
    \tilde S_\euc \equiv
    \int_{\tilde x_1}^{\tilde x_2} \du \tilde x
        \sqrt{2\tilde m (\tilde m\tilde V(\tilde x)-\tilde{E} )}
    .
\end{equation}
Accordingly, we also describe the Euclidean action $S_\euc$ as the \textit{WKB action.}

This result highlights two important implications of the semiclassical limit, which requires the integral to be larger than $\hbar$. First, it justifies neglecting corrections arising from the immediate vicinity of the turning points~\cite{Berry:1972na}. Second, it justifies neglecting the impact of the imaginary part of the energy on this analysis. On dimensional grounds, we can estimate the dimensionful prefactor $\tilde{\nu}$ in \cref{eq:width-estimate} as $\tilde{\nu}\sim \tilde E$. Therefore, in the semiclassical limit in which $\tilde \varphi/\hbar \gg 1$, we always have $|\tilde \Gamma| \ll |\tilde{E}|$. For completeness, we discuss how to move beyond the approximation of a strictly real energy in \cref{sec:exact-wkb-appendix}.

This analysis can be generalized in a straightforward way to our case of interest, namely, the tunneling rate out of the lowest-lying resonance in a radially symmetric potential. Anticipating a radially symmetric wave function $\psi=\psi(r)$ for the ground state, the time-independent Schr\"{o}dinger equation becomes
\begin{equation}
    -\frac{\hbar^2}{2\tilde m}\nabla^2\tilde \psi(\tilde r)
    +\tilde m \tilde V(\tilde r)\tilde\psi(\tilde r)
    =
    -\frac{\hbar^2}{2\tilde m}\left(
        \tilde \partial_{\tilde r}^2
        + \frac{2}{\tilde r}\tilde \partial_{\tilde r}
    \right)\tilde \psi(\tilde r)
    +\tilde m\tilde V(\tilde r) \tilde \psi(r)
    \overset{!}{=}
    \tilde{E}\psi(r)
    .
\end{equation}
Here and throughout the remainder of this work, we define $\gamma$ as the eigenvalue of the lowest-lying resonance, i.e., the state localized within the false vacuum basin with the lowest value of $E$. The additional term relative to \cref{eq:time-independent-schrodinger-1d} can be removed by introducing a rescaled wave function $u(r)\equiv r \psi (r)$, making $u(r)$ subject to \cref{eq:time-independent-schrodinger-1d}. Thus, \cref{eq:width-estimate} also applies to the lowest-lying resonance state in three dimensions, enabling a quick calculation of the rate of mass loss from the semiclassical energy of the dark matter particles.

\subsection{The instanton method}
\Cref{eq:width-estimate} can be established more directly using instanton methods, e.g., as first proposed in the seminal \refscite{Coleman:1977py,Callan:1977pt}, and more recently in the \textit{direct approach} proposed in \refscite{Andreassen:2016cff,Andreassen:2016cvx} and recently expanded to states other than a false vacuum in \refscite{Steingasser:2023gde,Steingasser:2024ikl,Lin:2025bjn,Lin:2025wgc,Barni:2026dhc}. For full details on the relevant derivations we refer to these works, see in particular~\cite{Lin:2025bjn,Barni:2026dhc}.

In this picture, the tunneling rate $\Gamma$ is defined through the time evolution of the probability $P_{\mathcal{F}}$ to find the particle in the basin $\mathcal{F}$ surrounding the false vacuum:
\begin{equation}
    \label{eq:PFdef}
    P_{\mathcal{F}}(t)\sim P_{\mathcal{F}}(0)\cdot e^{-\Gamma t}
    \;\iff\;
    \Gamma = - \frac{\dot{P}_{\mathcal{F}}(t)}{P_{\mathcal{F}}(t)}
    ,
    \qquad
    \text{where}\;P_{\mathcal{F}}(t) = \int_\mathcal{F}\du x\,|\psi(t,x)|^2
    .
\end{equation}
For a resonance state, this allows for the direct evaluation of the denominator in \cref{eq:PFdef} in the semiclassical regime, where $\Gamma \ll E$. The numerator can be expressed as a path integral, which can in turn be evaluated using a saddle point approximation. That is, schematically,
\begin{equation}
    \label{eq:PFfancy}
    \dot{P}_{\mathcal{F}} \sim \left|
        \int_0^t \du(\Delta t)\,
            \psi (x_1,t-\Delta t)
            \int_{x(0)=x_1}^{x(\Delta t)=x_2} \mathcal{D}x \ e^{iS[x]}
    \right|^2.
\end{equation}
The path integral describes the motion of a particle from $x_1$ to $x_2$, while the wave function accounts for the probability for the particle to be found in $x_1$ in the first place. The integral over $\Delta t$ accounts for paths that go from $x_1$ to $x_2$ with different tunneling times.

These integrals can then be evaluated through a combined saddle point approximation optimizing both the tunneling time $\Delta t$ and the chosen path. For tunneling out of excited states, this can be achieved using the so-called steadyon picture developed in \refscite{Steingasser:2023gde,Steingasser:2024ikl}. For the resonant states of interest for our purpose, this technique allows us to represent the dominant contribution to \cref{eq:PFfancy} in terms of easily calculable Euclidean-time quantities~\cite{Lin:2025bjn}:
\begin{align}\label{eq:PFLO}
    \dot{P}_{\mathcal{F}}\sim \left|  \psi (x_1) \right|^2
    \exp\left(- \frac{2}{\hbar} S_\euc[\bar{\tilde x}]+ \frac{2}{\hbar} \tilde{E} \Delta \tilde \tau\right).
\end{align}
The instanton $\bar{x}(\tau)$ denotes a solution to the Euclidean equation of motion
\begin{equation}
    \label{eq:eomEuc}
	0= \frac{\du^2}{\du\tau^2} \bar{x}(\tau)
        - V^\prime\bigl( \bar{x}(\tau)\bigr)
    ,
\end{equation}
subject to the boundary conditions
\begin{equation}
    \bar{x}(0) = x_1, \quad \bar{x}(\Delta \tau) = x_2,
    \quad \dot{\bar{x}}(0) = \dot{\bar{x}}(\Delta \tau) = 0
    ,
\end{equation}
as well as the \textit{crossing condition}, $x_1<\bar{x}(\tau)<x_s$ for $0<\tau<\Delta \tau$. Together, these conditions imply that the Euclidean-time energy of the instanton coincides with the energy of the resonance state $E$, as $V(x_1)=E$. This solution can be interpreted as the particle starting from $x_1$ at rest, and then rolling to $x_2$ in the inverted Euclidean potential. The Euclidean time $\Delta \tau$ is obtained as the time of the particle's first arrival at $x_2$.
Using conservation of the Euclidean-time energy, $E = mV(\bar{x}_I(\tau))- m \dot{\bar{x}}_I^2 (\tau)/2$, it is straightforward to demonstrate that \cref{eq:PFLO} is equivalent to the WKB result at leading order:\footnote{Note that this connection initially motivated the development of the instanton picture in \refscite{Coleman:1977py,Callan:1977pt}.}
\begin{multline}
    2S_\euc [\bar{x}] - 2E\Delta \tau
    = 2 m \int_0^{ \Delta \tau }\du\tau\left[
        \tfrac{1}{2} \dot{\bar{x}}^2
        + m V \bigl(\bar{x} (\tau)\bigr)
        - E
    \right]
    = 2m\int_{0}^{\Delta \tau} \du\tau
        \,\dot{\bar{x}}^2
    \\ 
    = 2m \int_{x_1}^{x_2} \du x
        \, \dot{\bar{x}}(\tau(x))
    = 2 \int_{x_1}^{x_2} \du x \left[
        2 m (m V(x)-E)
    \right]^{\frac{1}{2}}
    .
\end{multline}
As with the WKB result, this result also generalizes in a straightforward way to the three-dimensional setting: assuming a rotationally invariant potential, the instanton follows a straight contour~\cite{Barni:2026dhc}, $\bar{\bb x}(\tau) = \bar{r}(\tau)\bb e_{r}$, effectively reproducing the one-dimensional result under the replacement $\bar{x}(\tau)\to \bar{r}(\tau)$.

In general, the main advantages of the instanton picture relative to the WKB approximation are its relatively straightforward generalization to more complicated systems and calculation of next-to-leading order effects, which we defer to future work. For the remainder of this work, we use the WKB method to estimate tidal stripping rates.

\subsection{Application to tidal stripping}
An important subtlety of these semiclassical methods is that their application is usually limited to systems in which the relevant degrees of freedom are subject to an \textit{external} potential, whereas in the subhalo studied in this work the potential itself is (partially) generated by the wave function of the dark matter particles. In other words, applying these methods requires knowledge of the potential, which in turn depends on the wave function which we wish to analyze.

\refcite{Dave:2023egr} thus estimates the decay rate of the subhalo by neglecting all coupling between the gravitational potential $V$ and the tidal potential $V_\tidal$, taking $V_\eff(r) =V_0(r) + V_\tidal(r)$, where $V_0(r)$ is the potential found with $\omega = 0$, corresponding to a purely real energy eigenvalue $\gamma = E$. In this case, the exponent of $\Gamma$ simplifies to
\begin{equation}
    \label{eq:sigma-0}
    -\log \frac{\Gamma}{2} \sim S_\euc \equiv
        \int_{r_1}^{r_2}\du r\,\sqrt{
            2m\left(m V_{\eff}(r)-E\right)}
    = \int_{r_1}^{r_2}\du r\,\sqrt{
        2m \left(m V_0(r)
            -\tfrac{3}{2}m \omega^2r^2-E\right)}
    .
\end{equation}
We refer to this procedure as the \textit{perturbed stationary state method}. In \cref{sec:bohr-sommerfeld-appendix}, we use this method to show that $\log\Gamma \propto -1/\omega$ at leading order. Again, note that the energy eigenvalue $E$ is a property of the soliton, not of a single DM particle. Since the potential is sourced by the soliton itself, the system is effectively in dynamical equilibrium, from which the virial theorem implies that the energy is of the same order as the effective potential. Thus, $E$ cannot be neglected in \cref{eq:sigma-0}.

\refcite{Dave:2023egr} uses a simple form of this analysis as an order-of-magnitude validation of their other results. We have argued, however, that $E$ alone (rather than $\gamma$) is sufficient to reliably calculate $V_\eff$ due to the exponential suppression of the imaginary part. Since $E$ is also relatively easy to extract numerically, per \cref{sec:numerics}, this suggests the following procedure:
\begin{enumerate}
    \item Solve the system of \cref{eq:free-system} with real eigenvalue $\tilde{E}$ for the potential $\tilde{V}$, i.e., via
    \begin{equation}
        \label{eq:v-omega}
        \tilde{E}\tilde{\chi} =
        -\frac{\hbar^2}{2m}\tilde{\nabla}^2\tilde{\chi}
        +\tilde m \tilde{V}_{\eff}\tilde{\chi},
        \qquad
        \tilde\nabla^2 \tilde{V}
            = 4\pi G|\tilde{\chi}|^2.
    \end{equation}
    \item Using this potential, calculate $\Gamma$ in the semiclassical limit via
    \begin{equation}
        \label{eq:Gamma-dimensionful}
        \tilde\Gamma \simeq \tilde{\nu} \exp \left[
            - \frac{2}{\hbar}
            \int_{\tilde r_1}^{\tilde r_2} \du \tilde r \sqrt{
                2\tilde m\left(
                    \tilde m \tilde V_{\tilde\omega}(\tilde r)
                        -\tfrac{3}{2}\tilde m\tilde\omega^2\tilde r^2-\tilde{E}
                \right)
        }
        \right].
    \end{equation}
\item Compute the attempt frequency directly as the inverse of twice the transit time between the origin and the inner turning point: 
    \begin{equation}
        \label{eq:wkb-attempt-frequency}
        \nu = \frac{1}{4m} \left[
            \int_{0}^{r_1}
                \frac{\du r}{\sqrt{2m\bigl|
                    mV_\eff(r)-E\bigr|}}
        \right]^{-1}
        .
    \end{equation}
\end{enumerate}
Moving forward, we will use the algorithm of \cref{sec:numerics} in Step 1, and refer to the entire procedure as the \textit{BCI/WKB hybrid method}.

In Step 1, it is also possible to use the Bohr-Sommerfeld approximation of \cref{sec:bohr-sommerfeld} to obtain the real part of the eigenvalue, $E$, rather than solve the system of \cref{eq:v-omega} directly. Steps 2 and 3 can then be used unmodified to compute $\Gamma$. This provides a numerically simple prescription that also lends itself well to semianalytical heuristics, as we discuss further in \cref{sec:bohr-sommerfeld-appendix}. We refer to this procedure as the \textit{pure WKB method}.

\begin{figure}[t]
\centering
    \includegraphics[width=\linewidth]{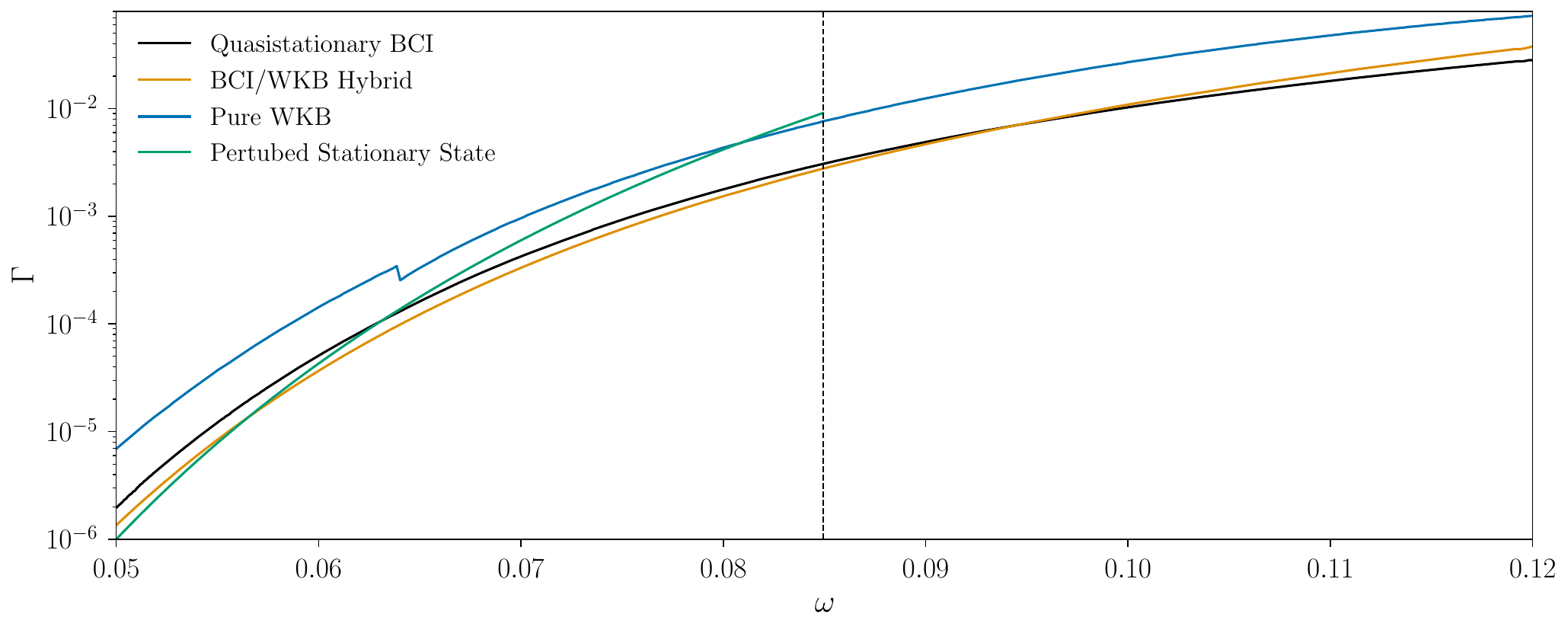}
    \caption{The imaginary part of $\gamma$ for a single-field soliton obtained through four different methods: \textbf{(1)} quasistationary BCI for the real and imaginary parts simultaneously (black); \textbf{(2)} quasistationary BCI for the real part and WKB tunneling approximation for the imaginary part (orange); \textbf{(3)} Bohr-Sommerfeld approximation for the real part and WKB for the imaginary part (blue); and \textbf{(4)} fixing the real part to that of the unperturbed $\omega=0$ solution and using WKB for the imaginary part (green). The unperturbed eigenvalue $E_0$ does not correspond to a metastable state in the presence of the tidal potential for $\omega \gtrsim 0.085$ (dashed vertical line).}
    \label{fig:decay-width-approximations}
\end{figure}

In \cref{fig:decay-width-approximations}, we compare the results for the imaginary part of $\gamma$ obtained through four different methods: direct quasistationary solution of the BCI system in \cref{eq:two-field-sys}; the BCI/WKB hybrid method; the pure WKB method; and the perturbed stationary state method. The quasistationary BCI and perturbed stationary state methods have appeared in the literature previously, whereas the BCI/WKB hybrid and pure WKB methods are newly developed in this work. \Cref{fig:decay-width-approximations} shows the single-field case, where the quasistationary BCI method is tractable and serves as a standard against which the other methods can be compared. We find that the BCI/WKB hybrid method shows excellent agreement with the quasistationary BCI results across a wide range of $\omega$ values. (The quasistationary BCI curve shown in the figure is the fit provided by \refcite{Du:2018qor}, which becomes unreliable at small $\omega$.) As anticipated, the perturbed stationary state method is effective at small $\omega$, but becomes unreliable at large $\omega$, where the perturbing tidal potential has a nonnegligible effect on the density profile and the resulting energy eigenvalue. For $\omega \gtrsim 0.085$, the energy eigenvalue obtained with $\omega=0$ corresponds to a scattering state rather than a metastable bound state in the perturbed potential, so the perturbed stationary state method incorrectly predicts a classically unbound state. The pure WKB method is accurate to within one order of magnitude across the $\omega$ values shown here, achieving these results without any explicit boundary condition at infinity. The discontinuity in the pure WKB decay rate at $\omega \approx 0.065$ is not a numerical artifact, but a consequence of the fact that continuity of $E$ does not guarantee continuity of $\Gamma$ under the WKB approximation: $\Gamma$ depends on the shape of the potential barrier, which can be discontinuous with $E$.

This comparison demonstrates that the BCI/WKB hybrid method is reliable over a wide range of $\omega$ values. In addition, the perturbed stationary state method gives accurate results for small $\omega$, while the very simple pure WKB method gives qualitatively valid results for a wide range of $\omega$ values. Each of these three methods is numerically far more tractable than the quasistationary BCI method. In the following section, we use the BCI/WKB hybrid method to evaluate physical decay rates for multifield halos.

\section{Tidal stripping of multifield halos}
\label{sec:multifield-lifetimes}
We now apply the results of the preceding sections to the decay of multifield solitons. The generalization of the results is straightforward: treating field $a$ is identical to the single field case. To treat field $b$, we must return to unscaled, dimensionful units with
\begin{equation}
    -\frac{\hbar^2}{2 \tilde{m}_b}\tilde\nabla^2\tilde\psi_b(r)
        +\tilde{m}_b\tilde V_\eff(r) \tilde\psi_b(r)=\tilde E_b\tilde\psi_b(r)
    .
\end{equation}
This produces $\tilde{p}_b(r)=\sqrt{2\tilde{m}_b(\tilde{m}_b\tilde{V}_\eff -\tilde E_b)}$. After converting to dimensionless, rescaled units, we obtain 
\begin{equation}
    \label{eq:wkb-exponent-b}
    \begin{cases}
    \displaystyle
        {S_\euc}_b(\omega) =
            \int_{r_1}^{r_2}\du r\,\sqrt{
                2\theta \left\{
                  \theta\left[V(r)-\tfrac{3}{2}\omega^2r^2\right]
                      -E_b(\omega)\right
                \}
            }
        ,
    \\[5mm]
        \displaystyle
        \nu_b = \frac{1}{4\theta} \left[
        \int_{0}^{r_1}
            \frac{\du r}{
                \sqrt{2\theta\,\Bigl|
                    \theta\left[V(r)-\tfrac{3}{2}\omega^2r^2\right]-E_b
                \Bigr|}
            }
        \right]^{-1}
        .
    \end{cases}
\end{equation}
The initial value problem is then specified by \cref{eq:two-field-sys,eq:two-field-bc}.

Our parametrization of a two-field soliton nominally involves seven parameters: $\tilde m_a$, $\tilde m_b$, $\tilde\omega$, $\tilde\rho_\core$, $\tilde\rho_\host$, $\tilde M_\soliton$, and the central field ratio $\eta$. In rescaled, dimensionless units, the only free parameters are $\omega$, $\theta$, and $\eta$. As with the single-field case, relationships between the parameters imply that a strict subset is enough to define a physical halo. In the case of a two-field halo, five parameters are sufficient, e.g. $\tilde{m}_a$, $\tilde\rho_\host$, $\omega$, $\theta$, and $\eta$. The first two are only necessary for scale-fixing, and are not needed to compute the dimensionless decay width $\Gamma$.

\begin{figure}
    \centering
    \includegraphics[width=\textwidth]{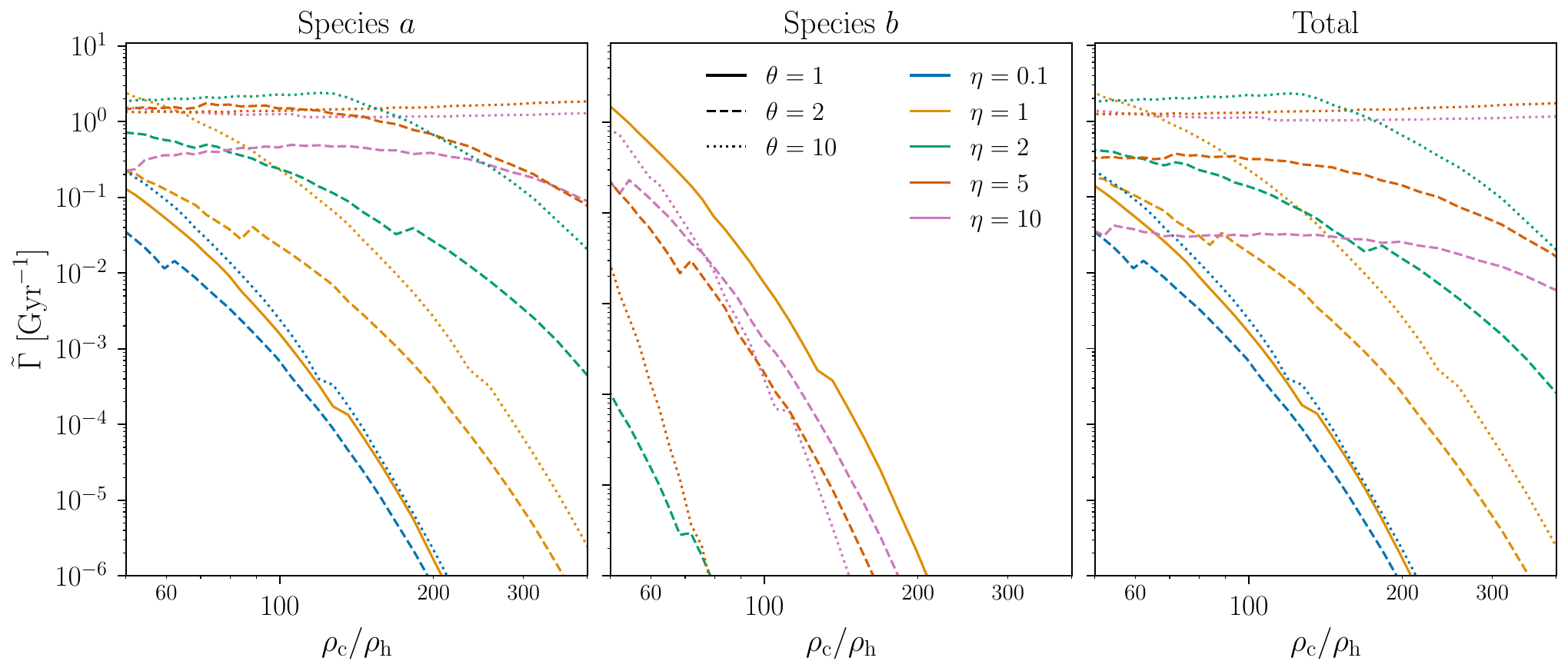}
    \caption{\ignorespaces
        Physical decay rates for the lighter field ($a$, left), heavier field ($b$, center), and overall halo (right) as a function of $\rho_\core/\rho_\host = 1/(3\omega^2)$, with fixed $\rho_\text{h}=\qty{e-3}{M_\odot/\parsec^3}$. The overall decay rate is computed as the soliton-mass--weighted average of the decay rates of the individual fields. Sharp features are a consequence of the numerical fitting procedure (see text). Masses are computed at $\omega = 0$, with mass ratios shown in \cref{fig:isolated_solitons}.
    }
    \label{fig:stability-bounds}
\end{figure}

With this extension of the semiclassical decay width calculation, we can finally use the BCI/WKB hybrid method to compute the decay rate of each field in a two-field halo. \Cref{fig:stability-bounds} shows dimensionful decay rates for a two-field halo as a function of $\rho_\core/\rho_\host$, fixing the host density $\tilde\rho_\host$ to \qty{e-3}{M_\odot/\parsec^3}. These curves are produced by fitting for the tidal response parameter $\alpha_i$ of \cref{eq:tidal-response}, and using the resulting values of $E_i$ as inputs to the WKB calculation. This produces some minor numerical artifacts, visible in \cref{fig:stability-bounds}, due to the fact that the values of $E_i$ produced by the fit deviate slightly from the exact values---and, as in \cref{fig:decay-width-approximations}, continuity of $E_i$ does not guarantee continuity of $\Gamma$. While these deviations are quite small, the system is very sensitive to deviations in the real part, as we have discussed in \cref{sec:numerics}. However, the results are qualitatively robust, and certainly sufficient to draw conclusions for the cosmological stability of two-field halos.

The solid orange curve in \cref{fig:stability-bounds} corresponds to $(\theta, \eta) = (1, 1)$, which is equivalent to the single-field solution. (The same is true at $\theta=1$ for any value of $\eta$.) This curve crosses $\tilde\Gamma \sim \qty{0.1}{\per\giga\year}$ at $\rho_\core/\rho_\host \sim 60$, becoming cosmologically unstable for lower values, consistent with previous calculations of the single-field soliton decay rate in halos with similar parameters~\cite{Hertzberg:2022vhk}. In the left and right panels, almost all other curves lie above this single-field curve, signaling higher net decay rates. These higher decay rates are driven entirely by the decay of the lighter field ($a$): the lighter field is much more susceptible to evaporation, having a larger Compton wavelength in the same potential as the heavy field. By contrast, note that the decay rates are uniformly suppressed in the middle panel, which shows the decay rate of the more centrally-concentrated heavy field ($b$).

The middle panel shows that as $\eta$ and $\theta$ both become large (e.g., when $\eta=\theta=10$), the decay rate for species $b$ is nearly as large as in the single-field solution. This is anticipated by the discussion in \cref{sec:two-field}: in this limit, the structure of the halo in the relevant range of radii is dominated by the heavy field, so the system returns to a single-field-like configuration, independent of the particle mass. A very similar result holds in the left panel for large $\theta$ and small $\eta$, where the light field dominates the halo. This accounts for the nonmonotonic behavior in $\eta$ at fixed $\theta$. Truly nontrivial effects on the decay rate appear when $\theta$ is different from but not very far from 1. This is exactly the parameter space where the limiting cases of \cref{sec:two-field} do not apply, and a nested-soliton description is inadequate. In this regime, the decay rate is suppressed for \textit{both} species. The greatest suppression occurs at small $\eta$ (dashed blue curves of \cref{fig:stability-bounds}): here the heavy field forms a subdominant but somewhat dense core that more tightly binds the halo. The suppression in the decay rate is mild, amounting to less than one order of magnitude for the cases we study. However, for such combinations, constraints from stability to tidal stripping are in fact alleviated relative to the single-field case: only halos with substantially lower $\rho_\core/\rho_\host$ would be destroyed on cosmological timescales.

Translating these results to the implications for a particular solitonic halo is nontrivial, because there is a complicated mapping between the dimensionless parameters $\{\omega,\theta,\eta\}$ and dimensionful, observable quantities such as $\tilde M_\soliton$ and $\tilde\omega$. In the single-field case, there is a very narrow range of soliton masses compatible with a given $\rho_\host$. The only relevant free parameter is $\omega$, and if $\omega$ becomes too large, the system admits no quasistationary resonance, i.e., there is no soliton solution. Accordingly, one can scan over $\omega$ and identify a range of accessible masses. One can then meaningfully evaluate the decay rate for each halo mass. This becomes much more complicated in the two-field case, where there are three free dimensionless parameters, and accordingly wider bands of accessible soliton masses.

\begin{figure}
    \centering
    \includegraphics[width=\textwidth]{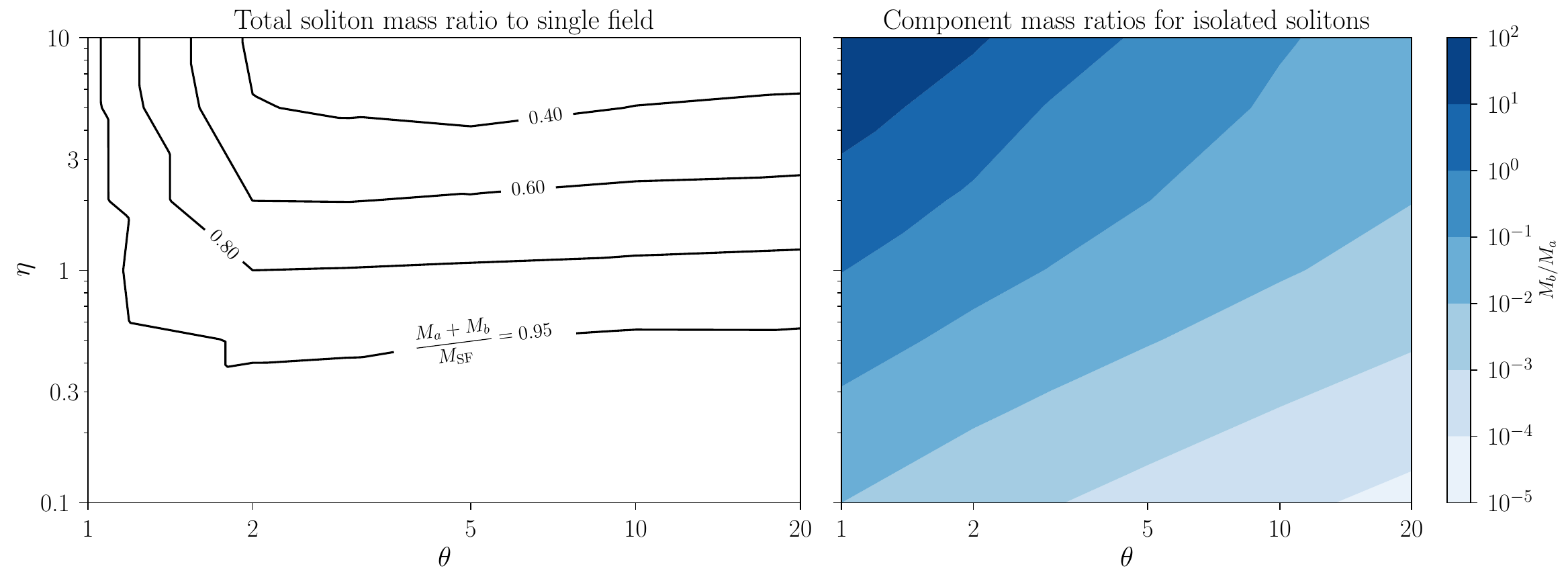}
    \caption{\ignorespaces
        Left: ratio of the total mass of the two-field soliton to that of a one-field soliton at fixed $\tilde\rho_c$ and at $\omega=0$ (i.e., in an isolated halo). The single-field mass is reproduced for $\eta \lesssim 1$ or $\theta\lesssim 2$.
        Right: ratio of component masses, $M_b/M_a$, at $\omega=0$.
    }
    \label{fig:isolated_solitons}
\end{figure}

The decay rates shown in \cref{fig:stability-bounds} have no explicit dependence on the dimensionful particle mass $\tilde m_a$ (or $\tilde m_b$). As discussed in \cref{sec:two-field}, the particle mass cancels out of dimensionless quantities. The implicit dependence on particle mass arises because of the relationship between particle mass and soliton mass, which in turn controls $\tilde\rho_\core$. For single-field solitons, the relationship between $\tilde m$ and $\tilde\rho_\core$ is well understood~\cite{Schive:2014dra}. However, recent simulations by \refcite{Huang:2022ffc} have shown that this relation does not apply to both components for arbitrary mass ratios. As such, we do not assume any such relation in computing our results, but rather evaluate the mass directly by integrating the profile. Note that while $\eta$ controls the central density ratio, it does not fully describe the mass ratio of the two soliton components. While we expect $M_i \propto \rho_{\core, i}$, and hence $M_b/M_a \propto \eta^2$, this mass ratio is also $\theta$-dependent. \Cref{fig:isolated_solitons} shows $M_b/M_a$ for our solutions as a function of $\theta$ and $\eta$ (right panel). We find that $M_b/M_a \sim \eta^2 / \theta$, consistent with findings in previous literature that the soliton mass scales with $\tilde m^{-1}$~\cite{Schive:2014hza,Huang:2022ffc}.

As anticipated in \cref{sec:sub:two-field-setup}, for many parameter points, the two soliton components have radically different decay lifetimes. For some of the curves in \cref{fig:stability-bounds} (e.g., $\theta=\eta=10$), the overall decay rate is dominated by the more massive soliton component, with particle mass $\tilde m_a$, which is unstable on cosmological timescales, while the less massive component is \textit{stable}. This leads to two different notions of stability for multifield halos. The first is the same as for single-field solitons: a soliton is unstable when the decay rate is large for both components, $\tilde\Gamma_a,\tilde\Gamma_b \gtrsim \qty{0.1}{\per\giga\year}$. The second is when $\tilde\Gamma_i \gtrsim \qty{0.1}{\per\giga\year}$ for just one of $i=a$ or $i=b$. In the second case, the two-field configuration is unstable, but rather than decay completely, the system will relax to a less massive single-field configuration. This behavior can also be understood in the tunneling analogy as an excited state that emerges from the superposition of two resonance states. For such states, a relatively large separation between the energies of these two states similarly manifests in the emergence of two distinct regimes~\cite{Lin:2025wgc,Lin:2025bjn}: for small times, the total decay rate is dominated by the larger contribution of the higher-energetic resonance state. As this corresponds to a quicker depletion of this state, its contribution eventually becomes subdominant, and the contribution of the lower-lying state becomes dominant.

Finally, we stress that while our results indicate whether particular combinations of parameters are cosmologically stable \textit{after formation,} they do not indicate whether or not such a halo will form in a realistic cosmology. In practice, $\theta$ is determined by the masses of the particle species themselves, but $\eta$ is a quantity that may vary from halo to halo. Computing the distribution of $\eta$ would require full cosmological simulations. Some suites of simulations have been conducted, and we refer the reader to \refscite{Huang:2022ffc,Luu:2023dmi} for further details.

\section{Conclusions}
\label{sec:conclusions}
In this work, we have developed a set of new techniques for studying the evaporation of ultralight DM subhalos induced by tidal potentials. While our primary aim has been the study of multifield halos, we have incidentally developed a framework for cleanly understanding and computing relevant quantities for single-field halos as well. In particular, we have shown in this work that \textit{solving the quasistationary eigenvalue problem to large radii is often unnecessary,} in both the single- and multifield cases. Since the quasistationary BCI method is typically ill-conditioned and not well suited to extract the imaginary part of the eigenvalue, it is valuable to have a simple and reasonably accurate alternative.

We began with the observation that the profile and effective potential resulting from solving the initial value problem at fixed $\gamma$ are relatively insensitive to the imaginary part, for reasons that are clear in hindsight: this must be true for any sufficiently narrow resonance (and thus sufficiently stationary state) to be of interest for cosmological observations. On this premise, we developed a new conceptual technique to solve for quasistationary states, first performing a much easier search for approximate \textit{stationary} solutions and then using WKB analysis to determine the imaginary part of the eigenvalue.

We further demonstrated that WKB analysis can be used to robustly estimate the eigenvalue entirely by means of the Bohr-Sommerfeld quantization condition. The quantization condition is enforced by a simple root-finding problem in one real variable, and only requires integrating the initial value problem to the inner turning point of the well. This eliminates the need to integrate to large $r$, and, with it, the extreme sensitivity to the real part of the eigenvalue. Furthermore, this obviates the need to perform a complex boundary condition matching procedure at large $r$.

Most importantly, however, WKB analysis gives clear intuition for the behavior of various features of the decay rate. In particular, we showed that for a significant portion of the parameter space of interest in $\omega$, the decay rate for a single field can be approximated via a single integral, and the scaling with $\omega$ is readily extracted. This would have been impossible in the context of the quasistationary BCI method used throughout the literature, since there are no analytical relations between the parameters and the behavior of the field at infinity, where the boundary condition is imposed.

Finally, we have applied our formalism to the study of multifield halos. Our numerical techniques make it relatively easy to study solitons with wide particle mass ratios across a wide range of tidal potential strengths $\omega$. We have computed physical decay rates for several classes of two-field halos, and for much of the parameter space, we have found that decay rates can be substantially enhanced. However, in the regime where the nested-soliton picture breaks down, i.e., where the two fields have an $\mathcal O(1)$ mass ratio and an $\mathcal O(1)$ core density ratio, we have found that the total decay rate can be somewhat \textit{suppressed}. This has significant implications for the applicability of ultralight DM constraints based on tidal stripping, suggesting that multifield scenarios with close mass ratios can alleviate the na\"ive single-field constraints---but that large mass ratios are not viable in much of the parameter space.

Recently, works such as \refscite{Luu:2023dmi,Luu:2024gnk} have also performed cosmological simulations of isolated ($\omega=0$) two-field solitons for $\eta\sim\mathcal{O}(0.1)$ and $\theta\sim\mathcal{O}(1)$, allowing for interactions between solitons. These interactions, in addition to the time dependence, make comparison between our results and theirs difficult. However, we are able to study large mass ratios, $\theta\sim\mathcal{O}(10)$, which have not yet been simulated. This regime is highly relevant for galactic phenomenology, since other works find that an order-of-magnitude mass ratio can explain the abundances of different types of dwarf galaxies~\cite{Pozo:2023zmx}.

We close by pointing out that although we have specialized in this work to the case of a tidal potential, our techniques are equally appropriate for the study of a wide class of perturbations for which the WKB method is valid, and would naturally extend e.g. to the case of a DM self-interaction, or mutual interactions between multiple fields in a DM halo. We defer such an analysis to future work.

\begin{acknowledgments}
The work of BVL is supported in part by Simons Investigator Award \#929255. JL is supported by the Moore Foundation Award 8342 and Harvard University institutional funding. TS acknowledges partial financial support from the Spanish Research Agency (Agencia Estatal de Investigaci\'on) through the grant IFT Centro de Excelencia Severo Ochoa No CEX2020-001007-S and PID2022-137127NBI00 funded by MCIN/AEI/10.13039/501100011033/FEDER, UE. This project has received funding/support from the European Union's Horizon 2020 research and innovation programme under the Marie Sk{\l}odowska-Curie Staff Exchange grant agreement No. 101086085-ASYMMETRY. JL thanks Cayla Dedrick and Rebecca Woody for their help in the early parts of this project.
\end{acknowledgments}

\appendix
\crefalias{section}{appendix}

\section{Bohr-Sommerfeld quantization and the stationary limit}
\label{sec:bohr-sommerfeld-appendix}
In \cref{sec:numerics}, we showed that the WKB approximation can give an estimate of the real part of the complex eigenvalue, in addition to the imaginary part discussed in \cref{sec:semiclassics}. In this appendix, we review the derivation and implications of the Bohr-Sommerfeld quantization condition, \cref{eq:bohr-sommerfeld-rule}. Our treatment closely follows \refscite{Barley:2025xro,griffiths_introduction_2018}. After this overview, we discuss the range of validity for the approximate treatments in \cref{sec:semiclassics}, and derive analytical results for small $\omega$. The results of this section are largely demonstrated in the single-field context, but the methods readily generalize to the multifield case.

In the case of a 1d potential, away from the classical turning points $r_i$, the wave function $\psi(\tilde r)$ is well described by \cref{eq:wkb-solution}, written in dimensionless variables as
\begin{equation}
    \psi(x)\approx
    \frac{1}{\sqrt{p(x)}}
    \begin{cases}
        \displaystyle
        C_1 \exp\left[i\int\du x\,p(x)\right]x_1
        + C_2 \exp\left[-i\int\du x\,p(x)\right]
        & 0<x<x_1,
        \\[4mm]
        \displaystyle
        C_3 \exp\left[-\int\du x\,|p(x)|\right] & x_1<x<x_2,
        \\[4mm]
        \displaystyle
        C_4 \exp\left[i\int\du x\,p(x)\right] & x_2<x.
    \end{cases}
\end{equation}
where $p(x) = \sqrt{2m[E-mV(x)]}$. For a 3d spherically-symmetric potential, the same form holds under the replacement $\psi(r)\rightarrow u(r) \equiv r\psi(r)$. However, as $x\rightarrow x_i$, the factor $1/\sqrt{p(x)}$ diverges, and so does $\psi(x)$. This nonphysical divergence is due to the failure of the WKB method at the turning points, and can be removed by connecting the solutions in \cref{eq:wkb-solution} with the appropriate \textit{connection functions} in the vicinity of the turning points.

Taking a turning point to lie at $x_i=0$ for brevity, near the turning point, we can write $V(x)$ as
\begin{equation}
    \label{eq:linearapprox}
    V(x)\approx E/m+V'(0)x.
\end{equation}
Plugging this back into \cref{eq:time-independent-schrodinger-1d}, and rearranging terms, we find 
\begin{equation}
    \partial_{z}^2\psi(z)= z\psi(z)
    ,
\end{equation}
where $z=\left(2m^2V'(0)\right)^{1/3} x$. This differential equation is solved by Airy functions, which serve as the connection functions for this case. The connection functions can then be used to bridge the two WKB solutions on either side of the classical turning point by matching at points far enough from $x_i$ for the WKB approximation to be valid, but close enough to $x_i$ that the approximation in \cref{eq:linearapprox} also remains valid. For a single turning point located at $x_1$, where $x<x_1$ is classically allowed and $x>x_1$ is not, $\psi(x)$ has the form 
\begin{equation}
    \label{eq:wkb-solutionconnected}
    \psi(x)\approx
    \begin{cases}
        \displaystyle
        \frac{C}{\sqrt{p(x)}}\sin\left[
            \int_{x}^{x_1}\du x'\,p(x')+\frac{\pi}{4}\right] & x < x_1
            ,
        \\[4mm]
        \displaystyle
        \frac{C}{\sqrt{|p(x)|}}\exp\left[
            -\int_{x_1}^{x}\du x'\,|p(x')|\right] & x_1<x
            .
    \end{cases}
\end{equation}
An analogous result holds for $u(r)$ in place of $\psi(r)$. In our case, boundary conditions fix $u(0) = 0$, which implies that the trigonometric function in the first line of \cref{eq:wkb-solutionconnected} must vanish when the integral is taken over the entire well, $0 < r < r_1$. Thus
\begin{equation}
    \label{eq:bohr-sommerfeld-rule-appendix}
    S_\euc \equiv
    \int_0^{r_1(E)}\du r\,\sqrt{
        2\left[E-V_\eff(E, r)\right]
    } =
    \left(n + \frac{3}{4}\right)\pi \text{~~for~} n\in\{0,1,2,\dotsc\}.
\end{equation}
This is the Bohr-Sommerfeld quantization condition for our system. The soliton ground state corresponds to the lowest energy $n=0$. The eigenvalue for the lowest-lying resonance is the unique value of $E$ for which $V_\eff$ and $r_1$ satisfy \cref{eq:bohr-sommerfeld-rule-appendix} with $n=0$.

Since $V$ and thus $V_\eff$ must be determined numerically for a given value of $\gamma$, it is necessary to solve \cref{eq:bohr-sommerfeld-rule-appendix} numerically. However, this numerical evaluation now takes the form of a simple 1-dimensional root-finding problem, and moreover, the domain of the integral in \cref{eq:bohr-sommerfeld-rule-appendix} is limited to the \textit{interior} of the well. The Bohr-Sommerfeld condition is completely insensitive to the large-$r$ behavior of the solution. Thus, in practice, solving for $\Re(\gamma)$ to satisfy the quantization condition can be implemented as a numerical shooting problem in which the differential system of \cref{eq:tidal-system} never needs to be solved for large $r$: the solution of the initial-value problem can be terminated as soon as the first turning point is reached.

Such a solution necessarily suffers from the inaccuracy associated with the WKB approximation, but we have already seen that WKB approximation is well behaved in the regime of interest in the context of the imaginary part of the eigenvalue, and, unsurprisingly, the same is often true for the real part. The blue curve in \cref{fig:decay-width-approximations} shows the values of $\Im(\gamma)$ obtained by the application of the WKB method for both the real and imaginary part: the Bohr-Sommerfeld quantization condition is used to determine $\Re(\gamma)$, which then determines $V_\eff$ and the turning points, which in turn fix $\Im(\gamma)$ via the WKB decay width. This pure WKB method agrees with the numerically-challenging quasistationary BCI computation (black curve) at the order-of-magnitude level.

In the stationary (small-$\omega$) limit, the Bohr-Sommerfeld method can compute the leading $\omega$-dependence of the real part of the eigenvalue, $E$. Recall that the single-field soliton profile at $\omega=0$ is well approximated by $\chi(r) = \exp(-ar^2)$, with $a \approx 0.1815$. This leads to a potential of the form
\begin{equation}
    V_\eff(r) \approx 
        \frac{4r\sqrt{a}
            - \sqrt{2\pi}\erf\left(r\sqrt{2a}\right)}
        {16r\left(\sqrt a\right)^3}
        - \frac32\omega^2r^2
    .
\end{equation}
Numerically imposing the Bohr-Sommerfeld condition of \cref{eq:bohr-sommerfeld-rule-appendix} at $\omega=0$ gives $E_0 \approx 0.6792$, comparable to the value $E_0 \approx 0.6495$ extracted by BCI methods in \cref{eq:single-field-tidal-response}. This is not quite an independent reproduction of the stationary eigenvalue, since the value of $a$ is taken from fitting to BCI results. But the virtue of this approach is that it becomes possible to incorporate small nonzero $\omega$ without solving the system again. Expanding the Bohr-Sommerfeld condition for small $\omega$, and taking $E = E_0 + \delta E$ for small $\delta E$, we obtain a condition of the form
\begin{equation}
    \int_0^{r_1}\du r\,\sqrt{2[E-V_\eff(r)]} \simeq
    \int_0^{r_1}\du r\,\left[
        \sqrt{2[E_0-V_0(r)]} + f(r)\,\delta E + \omega^2 g(r)
    \right] = \frac{3\pi}{4}
\end{equation}
for functions $f(r)$ and $g(r)$ given by
\begin{align}
    &f(r) = \frac{2a^{3/2}r\left[
        2\sqrt\pi\erf(r\sqrt{2a})-2r\sqrt{2a}(2-8aE_0)
    \right]}{
        \left[
            \sqrt{2\pi}\erf(r\sqrt{2a}) + 4r\sqrt{a}(4aE_0-1)
        \right]^2
    }\sqrt{16E_0-\frac4a + \frac{\sqrt{2\pi}\erf(r\sqrt{2a})}{ra^{3/2}}}
    ,
    \\
    &g(r) = \frac{
        3\sqrt\pi\erf(r\sqrt{2a})\,r^2 + 
        \left[
            24\sqrt2E_0a^{3/2}
            - 6\sqrt{2a}
        \right]r^3
    }{
        2\sqrt\pi\erf(r\sqrt{2a})-2r\sqrt{2a}(2-8aE_0)
    }\,f(r)
    .
\end{align}
The first term in the expanded integrand already satisfies the Bohr-Sommerfeld condition on its own, i.e., $\int_0^{r_1}\du r\,\sqrt{2[E_0-V_0(r)]} = 3\pi/4$. This means that
\begin{equation}
    \int_0^{r_1}\du r\,\left[
        f(r)\,\delta E + g(r)\,\omega^2
    \right] = 0
    ,
    \text{~~or~}
    \delta E = -\left(
        \frac{\int_0^{r_1}\du r\,g(r)}{\int_0^{r_1}\du r\,f(r)}
    \right)\omega^2
    .
\end{equation}
Evaluating numerically gives $\delta E \approx -6.71\omega^2$, or $E = 0.6792\bigl(1 - 9.88\omega^2 + \mathcal O(\omega^4)\bigr)$. This very nearly matches the small-$\omega$ behavior of the fit found in \cref{eq:tidal-response,eq:single-field-tidal-response}, which predicts $E \simeq 0.6495\bigl(1 - 9.12\omega^2 + \mathcal O(\omega^4)\bigr)$. In effect, the Bohr-Sommerfeld condition predicts $\alpha = 9.88$ for the single-field soliton, a result accurate to within 10\% and obtained without ever solving the system at nonzero $\omega$.

Similarly, at small $\omega$, the perturbed stationary state method discussed in \cref{sec:semiclassics} allows us to analytically study the scaling behavior of $\Gamma$. When $\omega$ is very small, it is perfectly legitimate to take $\Re(\gamma)$ from the $\omega=0$ case, and then use \cref{eq:sigma-0} to obtain $\Im(\gamma)$. This corresponds to the green curve in \cref{fig:decay-width-approximations}. Since $\Re(\gamma) = E$ is held constant in this computation as $\omega$ is increased, this fails catastrophically at large $\omega$: the soliton becomes unbound, meaning that $S_\euc(E) = 0$, which produces $\Gamma = -2\nu$. However, this only occurs for $\omega \gtrsim 0.09$, meaning that the perturbed stationary state method can be used to estimate decay rates for much of the parameter space of interest in this work.

The behavior of the perturbed stationary state computation is particularly simple to understand at very small $\omega$, and fully explains the behavior of the decay rate for single-field solitons in this regime. We can even obtain an analytical expression for the decay rate by approximating the shape of $V_0(r)$ via \cref{eq:long-range-potential-parameters}, which gives
\begin{equation}
    V_\eff(r) \approx V_\infty - \frac{A}{r} - \frac32\omega^2r^2,
    \text{~~where~}
    V_\infty = 1.377
    \text{~and~}
    A = 2.026
    .
\end{equation}
In the small-$\omega$ limit, the turning points can be found analytically as
\begin{equation}
    r_1 \simeq \frac{A}{V_\infty - E},
    \qquad
    r_2 \simeq \frac{A}{V_\infty - E}
        + \frac{1}{\omega}\sqrt{\tfrac23(V_\infty - E)}
    .
\end{equation}
Taking $E \simeq E_0 \approx 0.6495$ gives $r_1 \simeq 2.785$ and $r_2\simeq r_1 + 0.6964/\omega$. It follows that the decay width is approximated by
\begin{equation}
    \Gamma(\omega)\simeq
    \nu_0\exp\left\{
        -2
        \int_{\frac{A}{V_\infty - E_0}}^{
            \frac{A}{V_\infty - E_0}
                + \frac{1}{\omega}\sqrt{\tfrac23(V_\infty - E_0)}
        } \du r\,
        \sqrt{
            2\left(V_\infty - \frac{A}{r} - \tfrac32\omega^2r^2 - E_0\right)
        }
    \right\}
    ,
\end{equation}
where $\nu_0 \approx 0.069$ denotes the attempt frequency (\cref{eq:wkb-attempt-frequency}) for the $\omega = 0$ case. While the attempt frequency is in principle $\omega$-dependent, this dependence is mild compared to the exponential dependence in the small-$\omega$ regime. In the small-$\omega$ limit, the integral in the exponent can be directly evaluated, which gives
\begin{equation}
    \Gamma(\omega) \simeq
    \nu_0\exp\left(\frac{A}{2\sqrt{2(V_\infty-E_0)}}
        - \frac{10(V_\infty-E_0)}{3\sqrt3\,\omega}\right)
    \approx
    0.160\exp\left(-\frac{1.40}{\omega}\right)
    .
\end{equation}
One might expect from \cref{fig:single-field} that the $\log\Gamma$ should scale with $\omega^{-2}$. However, the $\omega^{-1}$ scaling can be readily validated numerically for $\omega \lesssim 0.03$ ($\rho_\core/\rho_\host \gtrsim 370$), beyond the edge of the figure. The physical reason for this scaling is simple: in the small-$\omega$ limit, the width of the potential barrier scales as $1/\omega$. For the slightly larger values of $\omega$ shown in \cref{fig:single-field}, numerical evaluation shows that indeed $\log\Gamma \sim -1/\omega^2$.

\section{Exact WKB method}
\label{sec:exact-wkb-appendix}
Our results for decay rates are valid at leading order in the semiclassical expansion. As discussed in \cref{sec:semiclassics}, this requires a significant exponential suppression of $\Gamma = -2\Im(\gamma)$ relative to $E = \Re(\gamma)$. We have argued that such a suppression is present for cosmologically-relevant scenarios. Nonetheless, in this appendix, we discuss how our results can be systematically improved when $\Gamma$ is non-negligible.

When $\Gamma \not\ll E$, the backreaction of $\Gamma$ on the potential and on the classical turning points becomes important. The turning points are defined by $V_\eff(r) - \gamma = 0$, which ordinarily assumes that $\gamma=E$ is purely real. In this case, the WKB action that governs the decay width can be written as
\begin{equation}
    \label{eq:wkb-exponent}
    S_\euc\bigl\{
        \omega,\,E;\,r_1[V_\eff,E],\,r_2[V_\eff,E]
    \bigr\}
    = \int_{r_1}^{r_2}\du r\,\sqrt{
        2\left\{V_\eff(r)-E\right\}
    }
    .
\end{equation}
But when $\gamma$ has a significant imaginary part, then clearly $V_\eff(r) - \gamma$ does not vanish at the original turning points $r_1$ and $r_2$, or indeed at \textit{any} real $r$. This is a general and well-known (if not widely-appreciated) phenomenon in any WKB tunneling analysis. The simple formula of \cref{eq:wkb-exponent} is usually applied to narrow resonances where the imaginary part is very small, but when $\Gamma$ is large, the shift in the turning points can be accounted for in the so-called \textit{exact WKB formalism} (see e.g. \refscite{voros1983return,delabaere1999resurgent,Sueishi:2020rug}).

In the exact WKB formalism, the eigenvalue in \cref{eq:wkb-exponent} is allowed to retain its imaginary part, and $r$ is analytically continued into the complex plane. The classical turning points $r_1$ and $r_2$ now become the complex-valued points at which both the real and imaginary parts of $V_\eff(r_i)-\gamma$ vanish, and for complex $\gamma$, we define
\begin{equation}
    \label{eq:exact-wkb-exponent}
    S_\euc\bigl\{
        \omega,\,\gamma;\,r_1(V_\eff,\gamma),\,r_2(V_\eff,\gamma)
    \bigr\}
    = \oint_{r_1}^{r_2}\du r\,\sqrt{
        2\left\{V_\eff(r)-\gamma\right\}
    }
    .
\end{equation}
One can integrate this WKB action on any contour between $r_1$ and $r_2$, but the analytic structure of the integrand requires the inclusion of particular contributions on crossing the branch cut along $V_\eff(r) = \gamma$. The branch structure of the integrand gives rise to what are known in the literature as \textit{Stokes lines,} which separate different asymptotic behaviors of the analytic continuation of the WKB action. The inclusion of these contributions is the generalization of the usual WKB connection formulas that one applies at the classical turning points: the turning points are simply the branch points at the end of the branch cuts in the complex-$r$ plane. Just as in the ordinary WKB approximation, appropriately counting contibutions across branch cuts (i.e., enforcing WKB connection formulas) is exactly equivalent to imposing a particular set of boundary conditions at infinity. This is ultimately the reason why WKB methods are able to estimate the full complex eigenvalues without ever integrating the solution to large $r$.

Now, since the exact WKB action $S_\euc(\gamma)$ depends on the imaginary part, we can write the following self-consistency relation:
\begin{equation}
    \label{eq:wkb-consistency}
    \nu(\omega,\gamma)\exp\Bigl[
        -2S_\euc\left(\omega,\gamma\right)
    \Bigr] = -2\Im(\gamma)
    .
\end{equation}
In principle, one could apply this relation \textit{iteratively} to fix $\Im(\gamma)$. That is, one could use the following procedure:
\begin{enumerate}
    \item Define an initial eigenvalue $\gamma^{(0)}$ from \cref{eq:wkb-exponent,eq:wkb-attempt-frequency}.
    \item Define the $i$th improved approximation as
    \begin{equation}
        \gamma^{(i)} \equiv \Re(\gamma^{(0)})
            - \frac12i\nu(\omega,\gamma^{(i-1)})\exp[
                -2S_\euc(\omega,\gamma^{(i-1)})]
        .
    \end{equation}
    \item Iterate until convergence of $\Im(\gamma^{(i)})$.
\end{enumerate}
The application of the WKB method in this procedure would be self-consistent, but note that $\Im(\gamma^{(i)})$ would not enter into the computation of $V_\eff$, so the computation would still not be entirely self-consistent. Nevertheless, we have implemented this iterative procedure and verified that for $\omega \ll 1$, the sequence of values $\Im(\gamma^{(i)})$ converges extremely quickly: $\Im(\gamma^{(0)})$ already provides an excellent approximation. This is exactly as expected in the semiclassical limit, where, as discussed in \cref{sec:semiclassics}, there is indeed an exponential hierarchy between $E$ and $\Gamma$.

\bibliographystyle{JHEP}
\bibliography{references}

\end{document}